\documentclass[prd,onecolumn,superscriptaddress,showpacs,nofootinbib,preprintnumbers]{revtex4}

\usepackage{amsmath}
\usepackage{amsfonts}
\usepackage{graphicx}
\usepackage{dcolumn}
\newcommand{\beq}{\begin{equation}}
\newcommand{\eeq}{\end{equation}}
\newcommand{\bea}{\begin{eqnarray}}
\newcommand{\eea}{\end{eqnarray}}
\newcommand{\beaa}{\begin{eqnarray*}}
\newcommand{\eeaa}{\end{eqnarray*}}

\begin{document}
\tolerance=5000


\title{Cosmic equation of state from combined angular diameter distances: Does the tension with luminosity distances exist?}
\author{Shuo Cao and Zong-Hong Zhu \footnote{zhuzh@bnu.edu.cn} }

\affiliation{
\medskip
Department of Astronomy, Beijing Normal University, Beijing 100875,
China }

\begin{abstract}

Using a relatively complete observational data concerning four angular diameter distance (ADD) measurements and 
combined SN+GRB observations
representing current luminosity distance (LD) data, this paper investigates the 
compatibility of these two cosmological distances considering three classes of
dark energy equation of state (EoS) reconstruction. In particular, we use strongly gravitationally lensed systems from various large
systematic gravitational lens surveys and galaxy clusters, which yield the Hubble constant independent ratio between two angular diameter
distances $D_{ls}/D_s$ data. Our results demonstrate that, with more general categories of standard ruler data,
ADD and LD data are compatible at $1\sigma$ level.
Secondly, we note that consistency between ADD and LD data 
is maintained irrespective of the EoS parameterizations: there is a good match between
the universally explored CPL model and other formulations of cosmic equation of state. Especially for the truncated GEoS model with $\beta=-2$,
the conclusions obtained with ADD and LD are almost the same.
Finally, statistical analysis of generalized dark energy equation of state
performed on four classes of ADD data provides stringent constraints on the EoS parameters $w_0$, $w_{\beta}$
and $\beta$, which suggest that dark energy was a subdominant component at early times. Moreover, the GEoS parametrization with $\beta\simeq 1$
seems to be a more favorable two-parameter model to characterize the cosmic equation of state, because the combined  angular diameter distance data (SGL+CBF+BAO+WMAP9) provide the best-fit value $\beta=0.751^{+0.465}_{-0.480}$.

\end{abstract}


\maketitle

\section{Introduction}

The discovery of the present acceleration of the cosmic expansion, which was firstly confirmed by the observations from the Type Ia supernova (SN Ia)
surveys \cite{Riess98,Perlmutter99} invokes a straightforward inclusion of an exotic source of matter with negative net pressure,
the so-called dark energy. There are, however, another theoretical approaches trying to explain cosmic acceleration
by modification of gravity at cosmological scales \citep{Nojiri03,Carroll04,Hu07}. So the nature of dark energy is still a mystery. Therefore, if
dark energy is responsible for the accelerating expansion of the Universe,
then it is necessary to study the  parameters effectively describing its properties, such as its density parameter and coefficients
in the cosmic equation of state (EoS): $w(z)=p_X/\rho_X$, where $\rho_X$ and $p_X$ are respectively
its density and pressure and $z$ is the redshift (See Refs\cite{Astier00,Weller01,Chevalier01,Linder03,Huterer07,Barboza09} for more parameterizations of
the $w(z)$ coefficient). It follows then that the dark energy density function in units of the critical density reads:
\begin{equation} \label{omx}
\Omega_X(z)\propto \exp[3\int^{z}_0(1+w(z')d\ln(1+z')].
\end{equation}

There are two direct probes of expansion history of the Universe, which can be tested observationally. One is the luminosity distance $D_L(z)$, and the other is the angular diameter distance $D_A(z)$. Theoretically, both of the expressions of the two cosmological distances are defined from the so-called coordinate distance
\begin{equation}
\label{eq:r} r = \frac{c}{H_0 \sqrt{|\Omega_k|}} {\rm
sinn}\left[\sqrt{|\Omega_k|} \int_{0}^{z}\frac{dz'}{E(z';\textbf{p})}\right],
\end{equation}
where $H_0$ is the Hubble constant, $c$ is the speed of light, $E(z; \textbf{p})=H/H_0$ is the expansion rate that has different forms with different
cosmological model parameters \textbf{p}, and $\Omega_k$ is the spatial curvature density parameter; $sinn(x)$ is $sinh(x)$ for $\Omega_k>0$, $x$ for
$\Omega_k=0$, and $sin(x)$ for $\Omega_k<0$, respectively. The angular diameter distance $D_A$ (hereafter ADD) and the luminosity distance $D_L$ (hereafter LD)
are simply related to the coordinate distance as
\begin{equation}
\label{eq:dA} D_A = r/(1+z)
\end{equation}
and
\begin{equation}
\label{eq:dL} D_L = r(1+z),
\end{equation}
At present there exist two classes of probes that may be used to observe the above two cosmological distances and thus equivalently $w(z)$ by searching this
sort of object at different redshifts. In order to measure the luminosity distance, we always turn to luminous sources of known (or standardizable) intrinsic luminosity in the universe, such as SN Ia and 
less accurate but more luminous Gamma Ray Bursts (GRB) in the role of "standard candles". 
On the other hand, in order to measure the angular diameter distance, we always turn to objects of known (or standardizable) comoving size acting as "standard rulers".
The most commonly used standard ruler in cosmology is the sound horizon at the epoch of last scattering, the scale of which can be measured either through BAO matter power spectrum (the bump in the galaxy correlation function due to baryon acoustic oscillations) or the CMB temperature spectrum. 
At low redshifts,
radio galaxies \cite{Gurvits99} and clusters of galaxies \cite{Boname06,Filippis05} may be used as standard rulers under certain assumptions.

The reason of contemplating ADDs and LDs separately is that these two kinds of distances are based on different principles. In the Euclidean space these distances coincide but in FRW spacetime i.e. in cosmological context they don't. Indeed they are related with each other by the so called Etherington principle:
$D_L = (1+z)^2 D_A$. Moreover, methodologies to measure ADD and LD are based on different physical principles. Therefore, these distinct classes of probes are prone to different systematics which makes them complementary and also motivates to make cosmological inferences on them separately.
There have been arguments based on the compatibility of results derived by using angular diameter distances and luminosity distances,
respectively, that a certain "tension" in the estimated values of EoS parameters can arise.
Specifically we will use the word "tension" to indicate that the EoS parameters values obtained with techniques using ADD and LD
differ from one another at least by $2\sigma$.
It was found in \cite{Lazkoz08} that systematic differences occur between cosmological parameters obtained from standard rulers (BAO and CMBR shift parameter R) and standard candles (SN Ia ESSENCE+SNLS+HST sample).
The shift in the best-fitted parameters inferred from standard candles and standard rulers was also noticed
and discussed in Ref.~\cite{Linder08,Biesiada10}. The compatibility between SNIa and BAO data was investigated in two different dark energy EoS reconstructions including the well known Chevalier-Polarski-Linder (CPL) model \cite{Escamilla11}.
Further papers have also noticed this disagreement in cosmographic studies using different probes.
Ref.~\cite{Biesiada11} made a joint ADD analysis with the strong gravitational lensing systems, the CMB acoustic peak location and
BAO data. While comparing the results from standard rulers  with those obtained from the Union2 SN compilation data \cite{Amanullah}, differences in central values of the best-fit cosmological parameters were also reported. More recently, the analysis performed by Planck mission team \cite{Planck1}  has revealed the tension between cosmological parameters inferred from Planck data and those from the SuperNova Legacy Survey (SNLS) 3-year data \cite{Conley11}, Union2.1 sample \cite{Suzuki12}, local measurements of $H_0$, and BAO constraints. The authors found the following results: (1) The Union2.1 best-fit is clearly compatible with Planck, especially for the value of $\Omega_m$
within the  $\Lambda$CDM model; (2) The combination of Planck and BAO measurements can give tight constraints on the cosmic equation of state $w\simeq -1$, which reflects the compatibility between BAO and CMB; (3) The SNLS SN sample, and direct measurements of $H_0$, are in tension with Planck at about the 2$\sigma$ level for the $\Lambda$CDM model. However, the mild tension seen between SNLS3 and BAO/CMB seems to have gone away with the recalibration efforts from the Sloan Digital Sky Survey/SNLS joint calibration.

Up to now, there exist several explanations of this tension or incompatibility between ADD and LD.
First of all, it may be just a statistical result produced by the limited amount
of observational data available \cite{Lazkoz08}. This has been noticed by many authors - e.g \cite{Zhai10},
who constrained the cosmological parameters of $\Lambda$CDM and XCDM cosmologies to examine the role of $H(z)$ and SN Ia data
in constraining cosmological models. In fact, the power of modern cosmology lies in building up consistency rather than in single, precise, crucial experiments.
In order to draw firm and robust conclusions about the consistency between ADD and LD, one will need to minimize statistical uncertainties by increasing
the depth and quality of observational data sets. Secondly, priors on the cosmological parameters including $\Omega_m$ and $\Omega_b$ may strongly influence the estimated values of the EoS parameters \cite{Sahni00,Lazkoz08}. However, by considering $1\sigma$ deviations of the matter fraction parameters for some priors,
there appears no tension among SN Ia, BAO and their combination (SN Ia+BAO) \cite{Escamilla11}. Thirdly, as long as compatibility is concerned, one cannot ignore the fact that tension may be brought by some caveats in the dark energy EoS parametrization.

In this context, it is clear that collection of more complete observational data concerning angular diameter distance measurements does play a
crucial role. 
The purpose of our paper is to show how the combination of most recent and
significantly improved cosmological observations concerning the cosmic microwave background (CMB) measurements \cite{Spergel07,Hinshaw09,Hinshaw12,Planck1,Wang13}, the baryonic acoustic oscillations (BAO) from large-scale structure considerations \cite{Eisenstein05}, and SN Ia \citep{Suzuki12} measurements can be used to probe the systematic differences between ADD and LD in the analysis on the cosmic equation of state.
In addition to previously studied probes we also use the strongly gravitationally lensed systems \cite{Cao12a},
the X-ray gas mass fraction of galaxy clusters \cite{Allen08}, and high redshift Gamma-Ray Bursts data \cite{Liang08,Wei10} to provide additional constraints on dark energy EoS. The idea of applying strong gravitational lensing systems
to probe the cosmic equation of state with the CPL parameterization was firstly discussed in Ref.~\cite{Grillo08} and also in more recent papers \cite{Biesiada10,Cao12a}.
In order to discuss the compatibility between LD and ADD in a general framework, more $w(z)$ parameterizations will be considered.

As for the calculating method, we choose to determine the best-fit values and the marginalized errors of each model parameter through the Markov Chain Monte Carlo (MCMC) method. The advantage of the MCMC method is that it allows for a simple inclusion of priors and a comprehensive study of the effects of systematic uncertainties. Our code is based on the publicly available CosmoMC package \cite{Lewis02}, which generated eight chains and stopped sampling when the worst e-values [the variance(mean)/mean(variance) of 1/2 chains] $R-1$ is of the order 0.01.

This paper is organized as follows. In Section~\ref{sec:method}, we briefly describe the methodology and observational samples for both angular diameter distances and luminosity distances.
Then, in Section~\ref{sec:model} we will introduce three classes of EoS parameterizations related to different dark energy models. We further present the results of constraining EoS parameters using MCMC method and test the compatibility between ADD and LD data in Section~\ref{sec:result}. In Section~\ref{sec:GEoS}, statistical analysis of the generalized dark energy equation of state involving
the four angular diameter distance tests are presented. Finally, the conclusions are summarized in Section~\ref{sec:conclusion}.

\section{The Method and the samples} \label{sec:method}

In order to probe dark energy models against observations, we consider four background tests which are directly related
to angular diameter distances: cluster baryonic fraction data (CBF), $D_{ls}/D_s$ data from SGL systems, baryonic acoustic oscillations (BAO),
and cosmic microwave background observations (CMB).
The first two tests are always considered as individual standard rulers while the other two
probes are treated as 
statistical standard rulers in cosmology. For the luminosity distances,
we choose to use supernova type Ia (SN Ia) and Gamma-ray bursts (GRB).

\subsection{Angular diameter distance observations}

The most direct angular size data were firstly derived from the observations of compact radio structures in
quasars and radio galaxies \cite{Gurvits99}. Then
it was found that we can also measure the angular diameter distances by using 
the Sunyaev-Zeldovich effect (SZE) together with X-ray emission of galaxy clusters \cite{Sunyaev72,Cavaliere78,Boname06,Filippis05}.
However, the 
constraining power of these ADD measurements is significantly affected by the large observational uncertainties. For example, the Bonamente sample \cite{Boname06} constrains $\Omega_m$ to about $\pm 0.7$ for the flat $\Lambda$CDM model, indicating that it is less than 1\% effective in constraining $\Omega_m$ compared with the latest BAO/CMB constraints. Therefore, in this paper we will incorporate two new datasets directly related to angular diameter distances: $D_{ls}/D_s$ data from strong gravitational lensing (SGL) systems and cluster baryonic fraction (CBF) data.

\subsubsection{$D_{ls}/D_s$ data from SGL systems}

As one of the successful predictions of General Relativity, strong gravitational lensing, which can generate multiple images of the background source at redshift $z_s$, is sensitive to angular distances between the source, the lens and the observer.
Combined image separation, redshift measurements and the spectroscopy of the lens can give us the ADDs.
Considering that the mass distribution of the elliptical galaxy acting as lens could be accurately described by the singular isothermal sphere (SIS) model \citep{Koopmans06,Koopmans09}, the Einstein radius in a SIS lens at redshift $z_l$ is
\begin{equation}\label{ringeq}
\theta_E=4 \pi \frac{D_A(z_l,z_s)}{D_A(0,z_s)}
                 \frac{\sigma_{SIS}^2}{c^2},
\end{equation}
where $\sigma_{SIS}$ can be identified with the central velocity dispersion. Correspondingly, the ratio of the angular diameter distances between lens and source and between observer and lens is rewritten as
\begin{equation}\label{ratio1}
\mathcal{D}^{obs}=\frac{D_{ls}}{D_{s}}\Big|_{obs}= \frac{c^2 \theta_E}{4\pi \sigma_{SIS}^2}
\end{equation}
Obviously, with the measured stellar velocity dispersion $\sigma_{0}$ from spectroscopy and the Einstein radius $\theta_E$ from image astrometry \citep{Treu06a,Treu06b,Grillo08},
SGL systems will provide us the third probe of ADD data in cosmology. However this is done indirectly --- through the ratio between two angular diameter distances.
We apply such a method to a combined gravitational lens data set, and selected 70 SGL systems from Sloan Lens ACS (SLACS) and Lens Structure and Dynamics
survey (LSD)
\citep{Cao12a}.

In this work we take a subsample including 64 galaxy-lens systems with the calculated distance between the lens and the source smaller than that
between the source and the observer, $D_{ds}/D_{s}<1$. Concerning the observational uncertainties both on the stellar velocity dispersion $\sigma_{0}$ and the Einstein radius $\theta_E$ \citep{Grillo08},
we obtain the corresponding uncertainty on the observational distance ratio $\mathcal{D}^{obs}$ calculated through the propagation of
uncertainty statics (See Table 1 of Ref.~\cite{Cao12a} for details).

Let us note here that the SIS model velocity dispersion $\sigma_{SIS}$ of the mass distribution and
the observed stellar velocity dispersion $\sigma_{0}$ may not be exactly equal and this is one of the systematics in this method. Based on the
observations in X-ray, it was argued that there is a strong indication that dark matter halos are dynamically hotter than the luminous stars \cite{White98}.
Therefore, a new parameter $f_E$ is included in our analysis to parameterize the relation between the stellar
velocity dispersion and the velocity dispersion in the form of \citep{Ofek03}
\begin{equation}
\sigma_{SIS}=f_{E}\sigma_{0}.
\label{f_E}
\end{equation}
In fact, the free parameter $f_E$ also reflects the effects of the rms error yielded by the assumption of the SIS model to relate
$\theta_E$ to the observed image separation $\Delta\theta$, as well as the decreasing of the typical image separations due to
the softened isothermal sphere potentials \citep{Narayan96}. For example, it was found that $f_E$ was in narrow range of 1 in our previous work \citep{Cao12a}.
So, it would be reasonable to assume it is a constant and include a 20\% uncertainty on the images separation due to
all these above factors, which is equivalent to the inclusion of a constant $f_E$ in the range $(0.8)^{1/2}<f_{E}<(1.2)^{1/2}$ \citep{Ofek03}.
Moreover, in order to obtain the constraint on the cosmological parameters of interest, the "nuisance" parameter $f_E$ is marginalized
by integrating off the full probability distribution function (PDF) \citep{Cao12a}.

For {\bf the} completeness, we also turn to SGL systems with clusters acting as lenses and galaxies acting as sources. This type of strong lensing can produce giant arcs around galaxy clusters with the observational arc position $\theta_{arc}$. If the hydrostatic isothermal spherical
symmetric $\beta$-model \citep{Cavaliere1976} can be used to describe the intracluster medium density profile, the Hubble constant independent
ratio can also be obtained
\begin{eqnarray}\label{ratio2}
\mathcal{D}^{obs}=\frac{\,\mu m_{p}c^{2}}{6
\pi}\frac{1}{k_{B}T_{X}\beta_{X}}\sqrt{
\theta_{t}\!^{2}+\theta_{c}\!^{2} }.
\end{eqnarray}
where $\beta_{X}$ and $\theta_{c}$ represent the slope and the core radius; $k_{B}$, $m_{p}$ and $\mu=0.6$ are the Boltzmann constant, the proton mass, and the mean molecular weight,
respectively \citep{Rosati2002}. The position of tangential critical
curve $\theta_{t}$ is usually deemed to be equal to the observational arc position $\theta_{arc}$.

Some authors applied such a method to SGL systems with both X-ray satellite observations as well as optical giant luminous arcs, and selected
10 lensing galaxy clusters in the redshift range $z=0.1-0.6$ \citep{Yu11}. The detailed information of X-ray galaxy clusters, $\beta$, $\theta_{c}$,
the redshift $z$, and the temperature $T_X$ are derived from the fitting results of Chandra, ROSAT, ASCA satellites and VIMOS-IFU survey \citep{Ota2004,Boname06,Covone06,Richard07}.
The final statistical sample of 10 SGL galaxy clusters with all the necessary parameters can be found in Ref.~\citep{Yu11}.

Therefore, in this work we take a sample with 74 observational $D_{ls}/D_s$ data points including 64 galaxy-lens systems and 10 cluster-lens systems,
which are selected from Table 1 of Ref.~\cite{Cao12a}, and the corresponding $\chi^{2}$
function is
\begin{equation}
\label{chiSGL}
\chi^2_{SGL}=\sum_{i}\frac{(\mathcal{D}_i^{th}(z_i;\mathrm{\textbf{p}})-\mathcal{D}_{i}^{obs})^{2}}{\sigma
_{i}^{2}}.
\end{equation}
where $\sigma _{\mathcal{D},i}^{2}$ denotes the $1\sigma$ error of the observational $\mathcal{D}_{i}^{obs}$.

\subsubsection{Cluster baryonic fraction data (CBF)}

Recently, the X-ray gas mass fraction of clusters, i.e., the cluster baryonic fraction versus redshift (CBF)
data from the Chandra satellite have become an effective probe at cosmological distances.
The matter content of the most massive galaxy clusters is expected to provide an almost fair
sample of the matter content of the universe. The ratio of baryonic-to-total mass in these clusters
should, therefore, closely match the ratio of the cosmological parameters $\Omega_b/\Omega_m$.
Because more than 80\% of clusters'
baryonic mass is in hot X-ray emitting intergalactic gas, a fair sample of
measurements of the cluster X-ray gas mass fraction (hereafter $f_{gas}$) from the detection of old, relaxed, rich clusters spanning some range of redshifts,
could provide an important source of ADD to probe the acceleration of the universe and therefore the cosmic equation of state \citep{Allen04,LaRoque06,Allen08,Samushia08,Ettori09}.

In this paper we will use the Chandra X-ray observations of 42 hot ($kT > 5 keV$), X-ray luminous, relaxed galaxy clusters in the
redshift range $z=0.05-1.1$ \citep{Allen08}, which have been shown to provide comparable constraints on dark energy to current SN Ia measurements \citep{Samushia08}.
Compared with the other astrophysical measurements, the CBF measures derived from X-ray observations are made within a given radius $r_{2500}$ for each cluster
($r_{2500}$ is the radius at which the mean enclosed mass density is 2500 times the critical density of the universe at the cluster-located redshift).
The $r_{2500}$ value for each cluster is determined directly from the Chandra data and these values may greatly differ from each other.
For the 42 galaxy clusters, the value of $r_{2500}$ ranges from $278^{+33}_{-25}h^{-1}_{70} \rm kpc$ (CL1415.2+3612 at $z=1.028$)
to $776^{+43}_{-31}h^{-1}_{70} \rm kpc$ (RXJ1347.5-1144 at $z=0.451$). From the Chandra data, reliable temperature measurements can also be made
at the outermost radii, which are generally well consistent with these $r_{2500}$ values. The detailed information of the 42 clusters (redshifts, $r_{2500}$ values, mean mass-weighted temperatures within $r_{2500}$ and the X-ray gas mass fractions within $r_{2500}$) can be found in Table 3 of Ref.~\cite{Allen08}.

We stress here that, in order to obtain constraints on the cosmological parameters of interest, fitting the reference $f_{gas}$ data set that
accounts for the expected variation in $f_{gas}$ is a more convenient method. The Allen sample \cite{Allen04} was used to work with the SCDM reference cosmology, however, as the Allen (2008) sample \cite{Allen08} clearly favours the $\Lambda$CDM over the SCDM cosmology. Therefore, in our analysis the $f_{\rm gas}$ measurements
in the reference cosmology is written as
\begin{equation}
  f_{\rm gas}(z;r_{2500}) =
A\, \Upsilon_{2500}\,\left(\frac{\Omega_{\rm b}}{\Omega_{\rm m}}\right)
\left(\frac{D^{\rm ref}_{\rm A}}{D_{\rm A}}\right)^{1.5},
\label{eq:fgas}
\end{equation}
where $\Upsilon_{2500}$ is the gas depletion parameter. The factor $A$, which is always very close to $1$, quantifies the shift in the angle subtended
as the cosmology of interest is varied. Compared with the previous studies neglecting the effect of $A$, we include it here to guarantee the accuracy of our analysis
\begin{equation}
A =  \left(\frac{r_{2500}^{\rm ref}}{r^{\rm }_{2500}}\right)^{\eta} \sim \left(\frac{[H(z)\,D_{\rm A}(z)]^{\rm }}{[H(z)\,D_{\rm A}(z)]^{\rm ref}}\right)^{\eta}\,.
\label{eq:angcosm}
\end{equation}
For the 0.7-1.2\,$r_{2500}$ shell, the slope factor is $\eta=0.214 \pm 0.022$ \citep{Allen08}. $D_{\rm A}$ and $D_{\rm A}^{\rm ref}$ are the angular diameter distances to the clusters computed
in the current model and reference flat $\Lambda$CDM cosmology with $\Omega_m=0.3$ and $h=0.7$ ($h$ is the reduced Hubble constant expressed as $H_{0}=100h  \rm km s^{-1} Mpc^{-1}$).

In order to account for other systematic uncertainties, Eq.~(\ref{eq:fgas}) is extended as \cite{Allen08}
\begin{equation}
\label{eq:fraction}
f_{\rm gas}(z)=\frac{KA\gamma b(z)}{1+s(z)}\left(\frac{\Omega_{\rm b}}{\Omega_{\rm m}}\right)\left(\frac{D_{\rm A}^{\rm ref}(z)}{D_{\rm A}(z)}\right)^{1.5}.
\end{equation}
Here $K$ is a calibration constant with a conservative $10\%$ Gaussian uncertainty $K=1.0\pm 0.1$ \citep{Allen08}.
The factor $\gamma$, with a uniform prior $1.1<\gamma<1.2$ \citep{Allen08}, models non-thermal pressure support in the clusters.
$b$ stands for the bias factor quantifying the difference between the baryon fraction in the cluster and the universe as a whole.
More specifically, this factor is modeled as $b=b_0(1+a_{\rm b}z)$ with priors $0.65<b_0<1.0$ and $-0.1<a_{\rm b}<0.1$ from gasdynamical simulation results \citep{Allen08}.
The parameter $s=s_0(1+a_{\rm s}z)$ models the baryon gas mass fraction in stars. We use the uniform prior with $-0.2<a_{\rm s}<0.2$ and the Gaussian prior with $s_0=0.16\pm0.05$ \citep{Allen08}.
The standard systematic uncertainties and priors on other parameters included in the Chandra CBF analysis can also be found in Ref.~\cite{Allen08}.
Like {]bf in} the case in SGL data, these CBF nuisance parameters are also marginalized over by multiplying the probability distribution function for each parameter and then integrating \citep{Ganga97,Allen08}.
Therefore, the resulting probability distribution function only depends on three variables: $\Omega_b$, $\Omega_m$ and the parameter \textbf{p} describing the cosmic equation of state.

In our analysis, in order to allow for systematics uncertainties, the rms fractional deviations in $K$, $\eta$, and $s_0$ are also added to the $\chi_{CBF}^2$
\begin{equation}
 \label{chiCBF}
\chi_{CBF}^2=\sum\limits_i
\frac{\big[f_{gas}^{th}(z_i;\textbf{p})-f_{gas}^{obs}(z_i)\big]^2}{\sigma_{i}^2}+\frac{(K-1)^2}{0.1^2}+\frac{(\eta-0.214)^2}{0.022^2}+\frac{(s_0-0.16)^2}{0.05^2}.
 \end{equation}
In the above expression, $f_{gas}^{obs}$ is the cluster gas mass fraction from observations and $\sigma_i$ is the total uncertainty of the CBF data for the $ith$ galaxy cluster.

\subsubsection{Baryonic acoustic oscillations (BAO)}

As it is well known, the baryonic acoustic oscillations (BAO) at recombination are expected to leave acoustic peaks in the power spectrum
of galaxies, which provides a standard ruler measuring the distance ratio
\begin{equation}
d_z=\frac{r_s(z_d)}{D_V(z_{\mathrm{BAO}})},
\end{equation}
where $r_s(z_d)$ stands for the co-moving sound horizon scale at recombination redshift $z_d$
\begin{equation}
r_s(z_{\ast}) ={H_0}^{-1}\int_{z_{\ast}}^{\infty}c_s(z)/E(z')dz'
\end{equation}
and the dilation scale $D_V$ is given by \citep{Eisenstein05}
\begin{equation} D_V(z_{\mathrm{BAO}})=\frac{1}{H_0}\big
[\frac{z_{\mathrm{BAO}}}{E(z_{\mathrm{BAO}})}\big(\int_0^{z_{\mathrm{BAO}}}\frac{dz}{E(z)}\big
)^2\big]^{1/3}~.
\end{equation}

Compared with previous works involving BAO as standard ruler \cite{Lazkoz08}, we use six precise measurements of the BAO distance
 ratio over a range of redshifts from $z=0.1$ to $z=0.7$ from the Sloan Digital Sky Survey (SDSS) data release 7 (DR7) \citep{Padmanabhan12},
 SDSS-III Baryon Oscillation Spectroscopic Survey (BOSS) \citep{Anderson12}, WiggleZ survey \citep{Blake12} and 6dFGS survey \citep{Beutler11}.
 we apply the maximum likelihood method using the data points with the best-fit values as \citep{Hinshaw12}
\begin{eqnarray}
\hspace{-.5cm}\bar{\bf{P}}_{\rm{BAO}} &=& \left(\begin{array}{c}
{\bar d_{0.10}} \\
{\bar d_{0.35}} \\
{\bar d_{0.57}} \\
{\bar d_{0.44}} \\
{\bar d_{0.60}} \\
{\bar d_{0.73}}\\
\end{array}
  \right)=
  \left(\begin{array}{c}
  0.336\pm0.015\\
  0.113\pm0.002\\
  0.073\pm0.001\\
  0.0916\pm0.0071\\
  0.0726\pm0.0034\\
  0.0592\pm0.0032\\
\end{array}
  \right).
 \end{eqnarray}
We find the contribution of BAO to the corresponding $\chi^2$ as
\begin{eqnarray}
\chi^2_{\mathrm{BAO}}=\Delta
\textbf{P}_{\mathrm{BAO}}^\mathrm{T}{\bf
C_{\mathrm{BAO}}}^{-1}\Delta\textbf{P}_{\mathrm{BAO}},
\end{eqnarray}
where $\Delta\bf{P_{\mathrm{BAO}}}=\bf{P_{\mathrm{BAO}}}-\bf{\bar{P}_{\mathrm{BAO}}}$, and ${\bf C_{\mathrm{BAO}}}^{-1}$ is the
corresponding inverse covariance matrix \cite{Hinshaw12}
\begin{eqnarray}
{\bf C_{\mathrm{BAO}}}^{-1}=\left(
\begin{array}{cccccc}
4444.4 &  0     &  0       & 0       & 0        & 0 \\
0      & 34.602 & 0        & 0       & 0        & 0 \\
0      &0       &20.661157 &0        & 0        & 0 \\
0      &0       &0         &24532.1  &-25137.7  &12099.1 \\
0      &0       &0         &-25137.7 &134598.4  &-64783.9 \\
0      &0       &0         &12099.1  & -64783.9 &128837:6 \\
\end{array}
\right)
\label{eq:normcovwmap}
\end{eqnarray}

\subsubsection{Cosmic microwave background observations (CMB)}

The second statistical standard ruler we use is the cosmic microwave background (CMB), which can provide the distance
at high redshift in order to determine the property of dark energy. Therefore, we implement the WMAP9 measurements of the derived quantities, such as the
angular scale of the sound horizon ($l_a$), the shift parameter ($R$), and the redshift of recombination ($z_{\ast}$). The angular
scale of the sound horizon at recombination can be parameterized as
\begin{equation}
l_a=\pi\frac{\Omega_\mathrm{k}^{-1/2}sinn[\Omega_\mathrm{k}^{1/2}\int_0^{z_{\ast}}\frac{dz}{E(z)}]/H_0}{r_s(z_{\ast})}.
\end{equation}
The commonly-used CMB shift parameter $R$ expresses as
\begin{equation}
R(z_\ast)=\frac{\sqrt{\Omega_{m}}}{\sqrt{|\Omega_{k}|}}{\rm
sinn}\left(\sqrt{|\Omega_{k}|}\int_0^{z_\ast}\frac{dz}{E(z)}\right).
\end{equation}
and the redshift of recombination $z\sim1089$ is more accurately written as
$z_{\ast}=1048[1+0.00124(\Omega_bh^2)^{-0.738}(1+g_{1}(\Omega_{\mathrm{m}}h^2)^{g_2})]$.
The values of relevant parameters $g_1$ and $g_2$ can be found in Ref.~\cite{Hu96}.
\begin{eqnarray}
g_1 &= &\frac{0.0783\, (\Omega_b h^2)^{-0.238}}
{1+39.5\, (\Omega_b h^2)^{0.763}}\\
g_2 &= &\frac{0.560}{1+21.1\, (\Omega_b h^2)^{1.81}}
\end{eqnarray}

For the flat prior, the 9-year WMAP data (WMAP9) measured best-fit values are \citep{Hinshaw12}
\begin{eqnarray}
\hspace{-.5cm}\bar{\textbf{P}}_{\rm{CMB}} &=& \left(\begin{array}{c}
{\bar l_a} \\
{\bar R}\\
{\bar z_{\ast}}\end{array}
  \right)=
  \left(\begin{array}{c}
302.40\\
1.7246\\
1090.88\end{array}
  \right).
 \end{eqnarray}
and we construct the contribution of CMB to the $\chi^2$ value as
\begin{eqnarray}
\chi^2_{\mathrm{CMB}}=\Delta
\textbf{P}_{\mathrm{CMB}}^\mathrm{T}{\bf
C_{\mathrm{CMB}}}^{-1}\Delta\textbf{P}_{\mathrm{CMB}},
\end{eqnarray}
with the corresponding inverse covariance matrix ${\bf C_{\mathrm{CMB}}}^{-1}$
\begin{eqnarray}
{\bf C_{\mathrm{CMB}}}^{-1}=\left(
\begin{array}{cccc}

   1.0000  &    0.5250  &   -0.4235  &   -0.4475    \\
  0.5250  &     1.0000  &   -0.6925  &   -0.8240    \\
 -0.4235  &   -0.6925  &     1.0000  &    0.6109    \\
 -0.4475 &   -0.8240  &    0.6109 &     1.0000  \\
\end{array}
\right)
\label{eq:normcov_planck}
\end{eqnarray}

In order to make a comparison with WMAP9, we will also use the distance priors from the Planck first data release \cite{Planck1,Wang13},
and examine their impact on the constraints of dark energy equation of state (See Section~\ref{sec:result}).

We will present a combined analysis of these above four tests to fit theoretical models to observational
data, i.e., the best-fit EoS parameters are obtained by minimizing
\begin{equation}
\label{chiADD}
\chi^{2}_{ADD}=\chi_{SGL}^{2}+\chi_{CBF}^{2}+\chi_{BAO}^{2} + \chi_{CMB}^{2}
\end{equation}

\subsection{Luminosity distance observations}

It is commonly believed that SN Ia can be calibrated as ``standard candles''.
SN Ia data do not provide the luminosity distance $D_L(z_i)$ directly, but rather the distance modulus defined as:
\begin{equation}
\mu_{th}(z_i)=m-M= 5 \log_{10} {D_L(z_i)/Mpc} +25.
\end{equation}
where $m$ and $M$ represent the apparent and absolute magnitude of a SN. In this paper, we use the latest Union2.1 compilation released by
the Supernova Cosmology Project (SCP) Collaboration consisting of 580 SN Ia data points \citep{Suzuki12}.
For the purpose of the likelihood calculations the $\chi^2$ value of the observed distance moduli can be calculated as follows:
\begin{equation}
\chi_{SN}^2=\sum_{i,j}[\mu(z_i)-\mu_{obs}(z_i)]C_{SN}^{-1}(z_i,z_j)[\mu(z_j)-\mu_{obs}(z_j)],
\label{chiSN1}
\end{equation}
where $\mu(z_i)$ is the theoretical value of the distance modulus, $\mu_{obs}(z_i)$ is the corresponding
observed value, and $C_{SN}(z_i,z_j)$ is the covariance matrix. Distance moduli $\mu_{obs}$ and the covariance matrix $C_{SN}$ are given in details in Ref.~\cite{Suzuki12} and can be found on the web site~\footnote{http://supernova.lbl.gov/Union/}.
There are two different covariance matrices corresponding to the cases with and without systematic errors. In this paper we will consider the case with systematic errors.
The nuisance parameter $H_0$ is marginalized with a flat prior, and Eq.~(\ref{chiSN1}) is rewritten as
\cite{Li11a}
\begin{equation}
\chi_{SN}^2=\sum_{i,j}\alpha_iC_{SN}^{-1}(z_i,z_j)\alpha_j-\frac{[\sum_{ij}\alpha_iC_{SN}^{-1}(z_i,z_j)-\ln10/5]^2}
 {\sum_{ij}C_{SN}^{-1}(z_i,z_j)}-2\ln\bigg(\frac{\ln10}{5}\sqrt{\frac{2\pi}{\sum_{ij}C_{SN}^{-1}(z_i,z_j)}}\bigg),
 \label{chiSN2}
\end{equation}
where $\alpha_i=\mu_{obs}(z_i)-25-5\log_{10}[H_0D_L(z_i)/c]$.

As an extension of previous works, we also add Gamma-Ray Bursts (GRBs) as complementary standard candles. Recently, Gamma-Ray Bursts (GRBs), which are the most luminous astrophysical events observable,
have been proposed as distance indicators at high redshift \cite{Schaefer03,Schaefer07,Wright07,Amati08,Daly08}.
The main advantage of GRBs over SN Ia is that they span a much greater redshift range, from low $z$ to $z>8$ \cite{Tanvir09}.
Moreover, comparing with SN, the high energy photons in the gamma-ray band are nearly unaffected by dust extinction.
Therefore, it may be rewarding to test the compatibility between LD and ADD with this newly obtained GRB data.
We use the "Hymnium" sample containing 59 data points, which were derived out of 109 long GRBs by applying the cosmology-independent luminosity
relation calibration method (the well-known Amati relation) \cite{Liang08,Wei10}. The "Hymnium" GRB sample is also given in terms of the distance modulus $\mu_{ obs}(z_i)$, which is included in our analysis by adding the following $\chi^2$ \cite{Pietro03}
\begin{equation}
\chi^2_{GRB}=\sum_{i}\frac{\alpha_i^2}{\sigma_{\mu{i}}^2}-\frac{(\sum_{i=1}^{59}\alpha_i/\sigma_{\mu{i}}^2-\ln 10/5)^2}{\sum_{i=1}^{59} 1/\sigma_{\mu{i}}^2}\nonumber\\
-2\ln\left(\frac{\ln 10}{5}\sqrt{\frac{2\pi}{\sum_{i=1}^{59}  1/\sigma_i^2}}\right),
\label{chiGRB}
\end{equation}

We will present a combined analysis of these above two tests to fit theoretical models to observational luminosity distance
data, i.e., the best-fit EoS parameters are obtained by minimizing
\begin{equation}
\label{chiADD}
\chi^{2}_{LD}=\chi_{SN}^{2}+\chi_{GRB}^{2}
\end{equation}

\section{ Cosmic equation of state tested} \label{sec:model}

In this paper, we assume three general classes of EoS parametrization for $w(z)$. To
derive the tightest possible constraints on the dark energy equation of state,
we assume a flat universe \citep{Hinshaw09}. This is due to the well known $\Omega_k$ - $w$ degeneracy.

\subsection{Dark energy with constant equation of state}

For the XCDM model, the equation of state parameter for dark energy is a constant $w$, and in such case this component is attributed to some sort of a evolving scalar field called quintessence or quintom \cite{Ratra88,Caldwell02}.
In a zero-curvature universe filled with ordinary pressureless dust matter (cold dark matter plus baryons), radiation and dark energy, the
Friedmann equation reads:
\begin{equation}
E^2(z; \textbf{p})= (\Omega_{b}+\Omega_{c})(1+z)^3+\Omega_{r}(1+z)^4+ \Omega_X(z).
\end{equation}
where $\Omega_b=(8\pi G\rho_b)/(3H^2_0)$ is the current baryonic matter component, $\Omega_{c}=(8\pi G\rho_{DM})/(3H^2_0)$ is the current dark matter component, the current radiation component $\Omega_{r}=(8\pi G\rho_r)/(3H^2_0)=4.1736\times 10^{-5}h^{-2}$ \cite{Komatsu09},
and the current dark energy component
\begin{equation} \label{w1}
\Omega_{X}(z)=(1-\Omega_{b}-\Omega_{c}-\Omega_{r})\times(1+z)^{3(1+w)}.
\end{equation}
Obviously, when flatness is assumed, it is a cosmological model with three parameters: $\textbf{p}=\{\Omega_{b}h^2,~\Omega_{c}h^2,~w\}$ and a nuisance parameter $H_0$.

\subsection{Dark energy with variable equation of state}

If we expect that $w$ coefficient vary in time, it could be an arbitrary function of the redshift, i.e. $w=w(z)$. In the following we will consider two parametrizations stemming from the first order Taylor expansions: in the scale factor $a(t)$ \citep{Chevalier01,Linder03} and in redshift $z$  \citep{Astier00,Weller01,Huterer07}. These are: the commonly used Chevalier-Polarski-Linder (CPL) $w(z) = w_0 + w_{P1}z/(1+z)$ and $w(z)=w_0+w_{P2}z$ respectively, where $w_0$ is the current value of the EoS parameter, and $w_{\rm{P}}$ (P = P1, P2) are free parameters quantifying the
time-dependence of the dark energy EoS. Note that the $\Lambda$CDM model can be always recovered by taking $w_0=-1$ and $w_{\rm{P}} = 0$.

In the universe filled with dark energy, ordinary pressureless dust matter
and radiation, the density fraction of dark energy can be expressed as
\begin{equation} \label{cpl2}
\Omega_X(z)=(1-\Omega_{b}-\Omega_{c}-\Omega_{r})\times(1+z)^{3(1+w_0+w_{P1})}\exp\left(-\frac{3w_{P1}z}{1+z}\right).
\end{equation}
in the CPL paramertization, and
\begin{equation} \label{cpl2}
\Omega_X(z)=(1-\Omega_{b}-\Omega_{c}-\Omega_{r})\times(1+z)^{3(1+w_0-w_{P2})}\exp\left(3w_{P2}z\right).
\end{equation}
for the second case of the equation of state of dark energy expanded up to the linear term in redshift.
There are four independent model parameters $\textbf{p}=\{\Omega_{b}h^2,~\Omega_{c}h^2,~w_0,~w_P\}$ and a nuisance parameter $H_0$ in this model.

\subsection{Dark energy with generalized equation of state (GEoS)}

Recently, a generalized EoS for dark energy was proposed \citep{Barboza09}
\begin{equation}
\label{Parametrization_a}
w(z)= w_0-w_{\beta}\frac{(1+z)^{-\beta}-1}{\beta}.
\end{equation}
From the above expression, it is straightforward to show that the above two variable EoS parameterizations are fully recovered when
$\beta\rightarrow +1$ and $\beta\rightarrow -1$, respectively. Obviously, the introduction of the new parameter $\beta$ is equivalent to put the EoS parameterizations (P1)-(P2) in a more general framework which admits a wider range of cosmological solutions ($w_{\beta}< 0$ or $w_{\beta}> 0$). For instance, some cases of interest relating the parameters $w_0$, $w_{\beta}$ and $\beta$ may be obtained as follows:
\begin{enumerate}
\item $\beta > 0$ ($w_{\beta}< 0$ or $w_{\beta}> 0$): at early times the dark energy is a subdominant component if $w_0 + w_{\beta}/\beta \leq 0$.
\item $\beta < 0$ and $w_{\beta} > 0$: at early times the dark energy always dominates over the other material components.
\item $\beta < 0$ and $w_{\beta} < 0$: at early times the dark energy density vanishes.
\end{enumerate}

For the generalized $w_{\beta}(z)$ model, the Friedmann equation for a spatially flat universe which contains only dust matter, radiation and dark energy, can be expressed as
\begin{eqnarray}
\label{FriedmannEquation_a}
\Omega_X(z)=(1-\Omega_{b}-\Omega_{c}-\Omega_{r})\times(1+z)^{3(1+w_0+w_{\beta}/\beta)}\exp\Big[\frac{3w_{\beta}}{\beta}\big(\frac{(1+z)^{-\beta}-1}{\beta}\big) \Big]\;,
\end{eqnarray}
where $\textbf{p}=\{\Omega_{b}h^2,~\Omega_{c}h^2,~w_0,~w_{\beta},~\beta,~H_0\}$.

\begin{table}
\caption{\label{result} Fits to different EoS models from combined ADD and LD data.}
\begin{center}
{\footnotesize
\begin{tabular}{cccccccc}
\hline\hline
EoS parametrization          & SGL+CBF+BAO+WMAP9   &  SGL+CBF+BAO+Planck    &  SN+GRB     \\
\hline

$w=const$    &  $\Omega_m= 0.302\pm0.024$  &  $\Omega_m= 0.300\pm0.025$  &  $\Omega_m=0.308^{+0.112}_{-0.226}$\\
             &   $w=-0.958\pm0.166$   &  $w=-1.110\pm0.141$     & $w=-0.972\pm0.450$   \\

  \hline

$w=w_0+w_{P1}\frac{z}{1+z}$  &   $\Omega_m=0.292\pm0.035$ &   $\Omega_m=0.305\pm0.032$  &  $\Box$ \\
                   &   $w_0=-1.050\pm0.375$  &   $w_0=-0.960\pm 0.410$    &  $\Box$    \\
                   &   $w_{P1}=0.440\pm1.250$ &   $w_{P1}=-0.600\pm 1.250$  &  $\Box$ \\

With WMAP9 priors on matter densities    &   $w_0=-1.050\pm 0.340$   &  $\Box$     &  $w_0=-0.975\pm 0.350$      \\
                                      &   $w_{P1}=0.400\pm0.800$ &  $\Box$     &   $w_{P1}=0.025\pm1.425$   \\

With Planck priors on matter densities    &  $\Box$    &  $w_0=-1.000\pm0.382$     &  $w_0=-0.950\pm0.365$      \\
                                       & $\Box$   &  $w_{P1}=-0.240\pm 1.120$     &   $w_{P1}=-0.240\pm1.645$   \\

\hline

$w=w_0+w_{P2}z$     &   $\Omega_m=0.305\pm0.026$ &   $\Omega_m=0.308\pm0.026$  &  $\Box$ \\
                    &   $w_0=-0.900\pm0.200$  &   $w_0=-0.900\pm0.250$    & $\Box$     \\
                    &   $w_{P2}=-0.100^{+0.150}_{-0.500}$ &   $w_{P2}=-0.250^{+0.250}_{-0.680}$  & $\Box$  \\

With WMAP9 priors on matter densities    &   $w_0=-0.900^{+0.198}_{-0.084}$   &  $\Box$     &  $w_0=-0.875^{+0.195}_{-0.135}$      \\
                   &   $w_{P2}=-0.07^{+0.100}_{-0.195}$ &  $\Box$     &   $w_{P2}=-0.100^{+0.110}_{-0.740}$   \\

With Planck priors on matter densities    &  $\Box$    &  $w_0=-0.930\pm0.206$     &  $w_0=-0.880\pm0.190$      \\
                                        & $\Box$   &  $w_{P2}=-0.085^{+0.118}_{-0.475}$     &   $w_{P2}=-0.175^{+0.175}_{-0.825}$    \\

\hline

GEoS ($\beta=+2$)    &   $\Omega_m=0.303\pm0.033$ &   $\Omega_m=0.310\pm0.033$  &  $\Box$ \\
                     &   $w_0=-0.965\pm0.347$  &   $w_0=-0.850\pm0.460$    &   $\Box$   \\
                     &   $w_{P3}=-0.340\pm1.015$ &   $w_{P3}=-0.800\pm1.900$  & $\Box$  \\

With WMAP9 priors on matter densities    &   $w_0=-0.720\pm0.245$   &  $\Box$     &  $w_0=-0.860\pm0.250$      \\
                   &   $w_{P3}=-0.730\pm0.715$ &  $\Box$     &   $w_{P3}=-0.705^{+0.805}_{-1.495}$   \\

With Planck priors on matter densities    &  $\Box$    &  $w_0=-0.880\pm0.335$     &  $w_0=-0.820\pm0.310$      \\
                   &  $\Box$  &  $w_{P3}=-0.400\pm1.125$     &   $w_{P3}=-0.500^{+0.710}_{-2.190}$    \\

\hline

GEoS ($\beta=-2$)    &   $\Omega_m=0.277\pm0.018$ &   $\Omega_m=0.307\pm0.019$  &  $\Box$ \\
                    &   $w_0=-1.055^{+0.053}_{-0.133}$  &   $w_0=1.061^{+0.070}_{-0.145}$    &  $\Box$    \\
                    &   $w_{P4}=-0.015^{+0.030}_{-0.075}$  &   $w_{P4}=-0.024^{+0.042}_{-0.085}$  & $\Box$  \\

With WMAP9 priors on matter densities   &   $w_0=-1.015^{+0.020}_{-0.090}$   &  $\Box$     &  $w_0=-1.022\pm0.023$      \\
                   &   $w_{P4}=-0.004^{+0.007}_{-0.022}$ &  $\Box$     &   $w_{P4}=-0.009^{+0.011}_{-0.035}$   \\

With Planck priors on matter densities  &  $\Box$    &  $w_0=-1.050^{+0.048}_{-0.092}$     &  $w_0=-1.051\pm 0.041$      \\
                   & $\Box$  &  $w_{P4}=-0.015^{+0.019}_{-0.064}$     &   $w_{P4}=-0.019^{+0.019}_{-0.075}$   \\

\hline\hline
\end{tabular} }
\end{center}
\end{table}

\section{Comparing data sets and testing their compatibility} \label{sec:result}

Next we shall test the compatibility between the angular diameter distance (SGL+CBF+BAO+CMB)
and the luminosity distance (SN+GRB). To compare these two data sets we perform fits of different
cosmological scenarios and obtain the constraint results displayed in Table~\ref{result}.
We take the following criterion of compatibility: if the confidence contours of $w_0$ and $w_P$ from ADD data shift away from the corresponding confidence contours from LD data, in a way that $1\sigma$ contours do not overlap, we call it a tension, otherwise we conclude that results are compatible.
Results show that the combination with angular diameter
distance data reveals no tension in all cases, though it also shows different features with respect to the previous references.
In fitting cosmological parameters within evolving EoS scenarios by using standard candles, we use the priors on the $\Omega_m$ either from WMAP9 or from Planck. The reason is that we want to obtain more stringent constraints on EoS parameters since the size of the confidence region is at the core of our study.
Fitting the $\Omega_m$ would lead both to very uncertain constraints on the density parameter itself and also would inflate the confidence regions for the EoS coefficients. We have verified this numerically. Although priors always influence the analysis (especially when obtained assuming a certain cosmological model), conclusions concerning compatibility of ADD and LD data should not be affected due to the lack of constraint on $\Omega_m$ for the models we are using.

\subsection{Dark energy with constant equation of state}

\begin{figure}
\begin{center}
\includegraphics[angle=0,width=85mm]{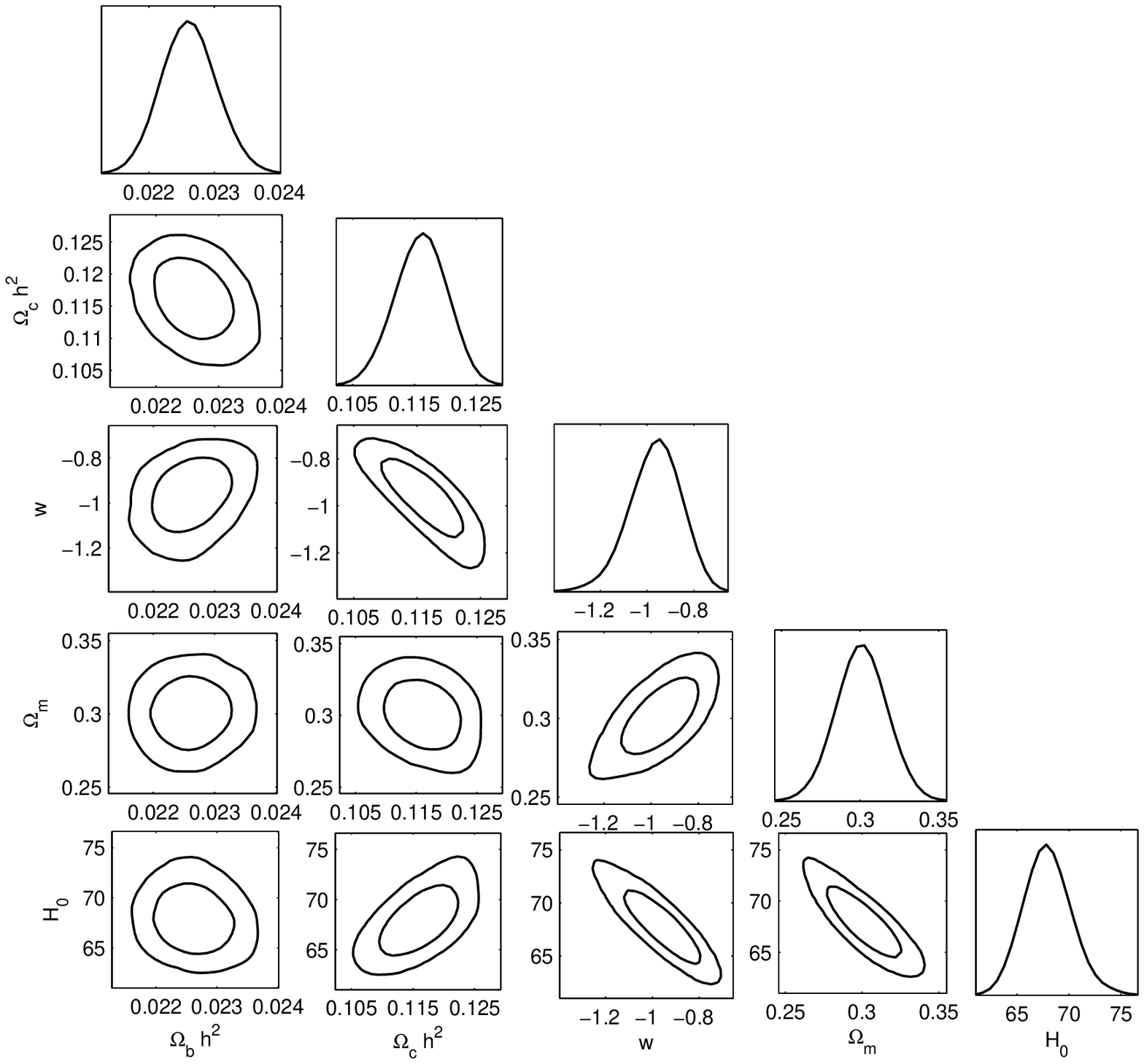}
\includegraphics[angle=0,width=75mm]{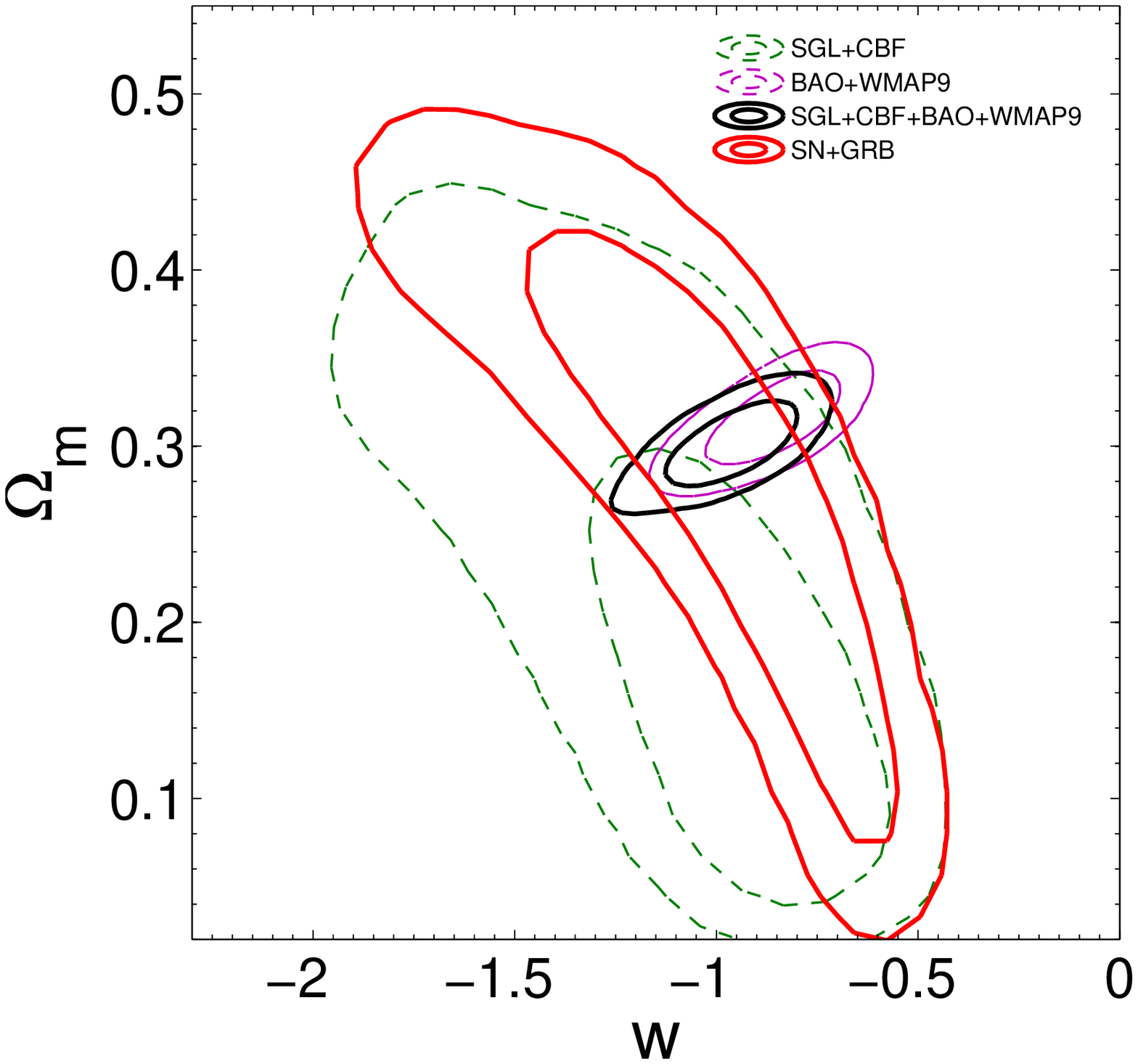}\includegraphics[angle=0,width=75mm]{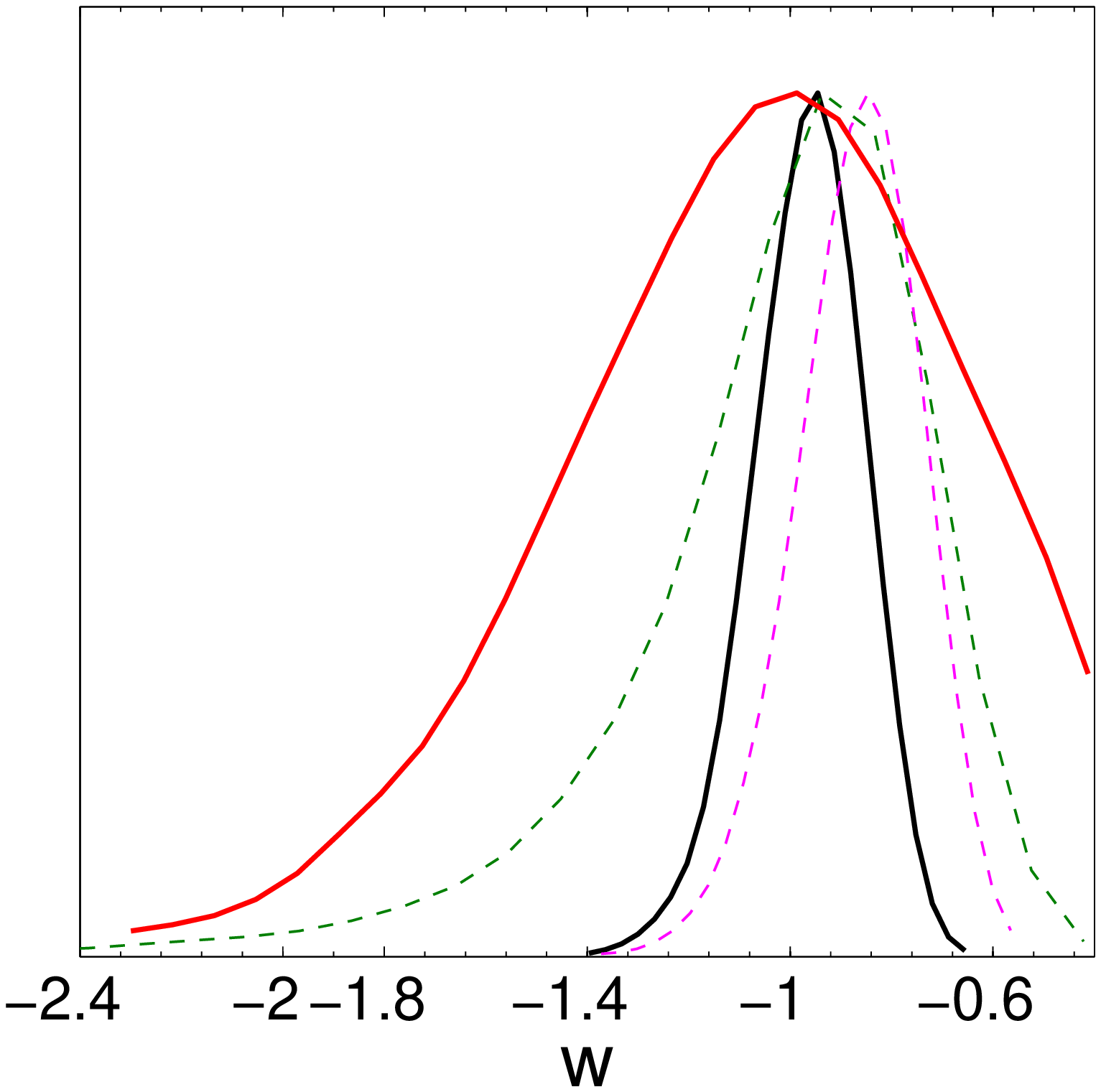}
\end{center}
\caption{ The 2-D regions and 1-D marginalized distribution with the
1-$\sigma$ and 2-$\sigma$ contours for parameters of the XCDM model from SGL+CBF+BAO+WMAP9 (black line), SN+GRB (red line), SGL+CBF(green line), and BAO+WMAP9 (purple line) respectively.
\label{XCDM}}
\end{figure}

By fitting the XCDM model to the above combined standard rulers, we get $\Omega_bh^2=0.0227\pm0.0009$, $\Omega_{c}h^2=0.1158\pm0.0075$, $w=-0.958\pm0.166$, and $H_0=67.50\pm3.55 \rm km s^{-1} Mpc^{-1}$.
After marginalizing over $H_0$, we obtain the dust matter density parameter $\Omega_m= 0.302\pm0.024$.
As shown in Table~\ref{result} and Fig.~\ref{XCDM}, we find the standard ruler data give strong preference for the flat quintessence dark energy model,
which is quite different from the results supporting a best fit that crosses the phantom divide line $w=-1$ \cite{Lazkoz08}. These results are also
consistent with those obtained from different sets of standard candle probes including Union1 SN Ia compilation \cite{Kowalski08}.

For comparison, in the following analysis our attention will be paid to the fits with the combined luminosity distance data.
The joint contour plot of $\Omega_{m}$ and $w$ (corresponding to 68.3\%and 95.4\% CL) for standard rulers and standard candles are shown
in Fig.~\ref{XCDM}. Black lines indicate the results from combined ADD data and red lines are from the combined LD results. In order to examine
the compatibility between the ADD and LD data in constraining the cosmic EoS parameter, the one-dimensional probability distribution function (PDF) of $w$
is also plotted in Fig.~\ref{XCDM}. It is obvious that the constraints of $w$ using the two data combinations are consistent with each other.
On the one hand, compared with the previous fitting results \cite{Lazkoz08,Biesiada11}, there is nearly no tension when
comparing the best fits from ADD data and LD data, because the separation is smaller than $1\sigma$. On the other hand, the result tells us
that ADD and LD data are especially concordant in constraining $\Omega_m$: the 1$\sigma$ confidence region of $\Omega_m$ achieved from SGL+CBF+BAO+WMAP9 is
$\Omega_m= 0.302\pm0.024$, while SN+GRB suggest $\Omega_m=0.308^{+0.112}_{-0.226}$.

In order to illustrate how much weight individual standard rulers and statistical standard rulers have in our analysis \cite{Gong10a,Gong10b},
we present the constraint results from the individual standard ruler joint analysis (SGL+CBF) in Fig.~\ref{XCDM}. We find the constraint
result of the EoS parameter $w=-0.948\pm0.375$ agrees very well with the SGL+CBF+BAO+WMAP9 constraint and is clearly in agreement with that obtained from LD data at $1\sigma$.
Meanwhile, in the case where the BAO+WMAP9 data are not combined with SGL+CBF, the data contours and one-dimensional marginalized probability distribution are obviously shifted, which demonstrates the non-negligible effect of the independent standard ruler data on model constraints.
On the other hand, we also note the other two statistical standard rulers, BAO and CMB, which are always taken as priors in the treatment and combined with other data, can tightly constrain the matter density $\Omega_{m}$ \cite{Gong12}. Therefore, the agreement between the independent standard ruler (SGL+CBF) and statistical standard ruler (BAO+CMB) constraints is reassuring and motivates the combination of these data sets.

\subsection{Dark energy with variable equation of state}

\begin{figure}
\begin{center}
\includegraphics[angle=0,width=85mm]{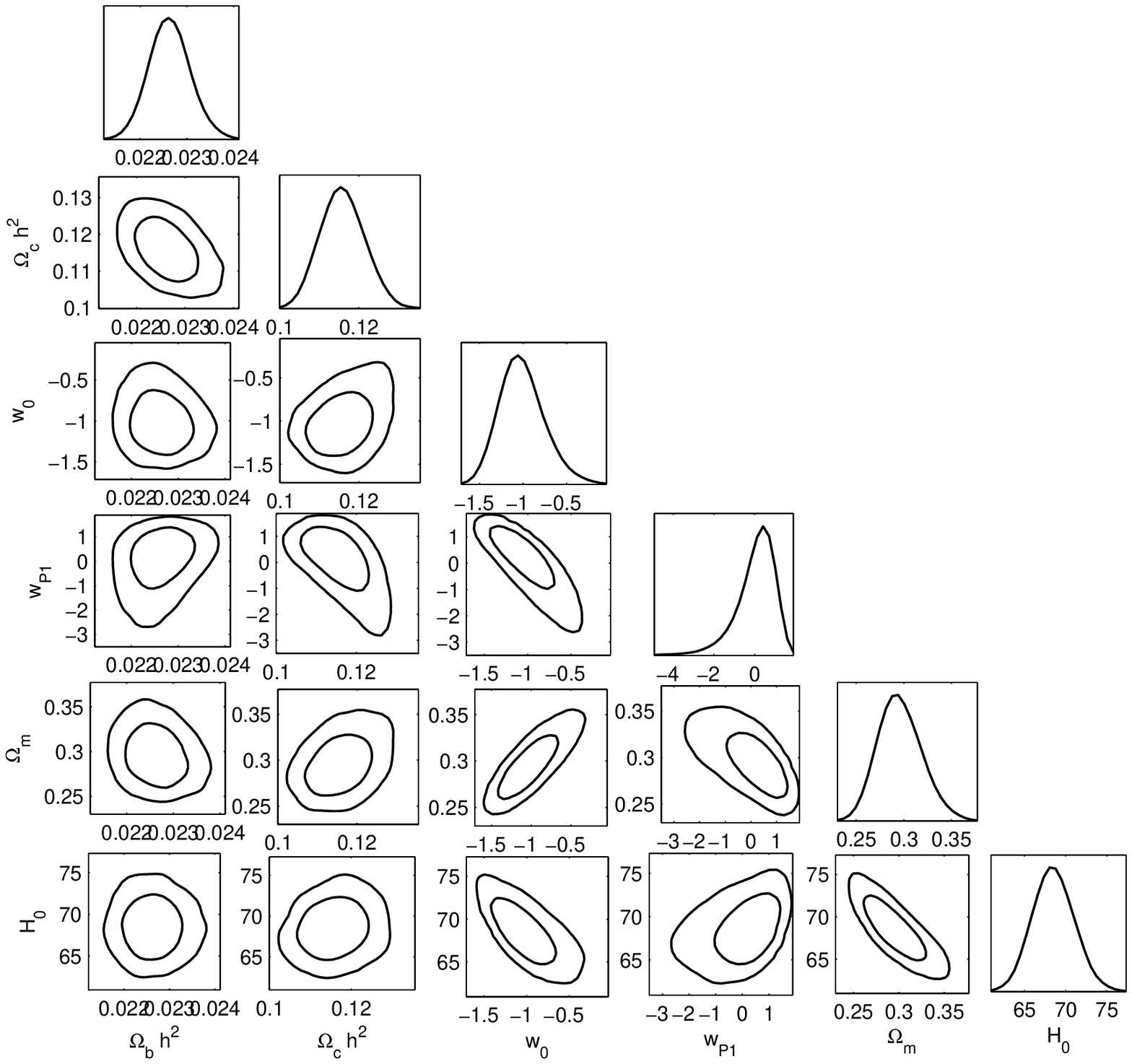}
\includegraphics[angle=0,width=75mm]{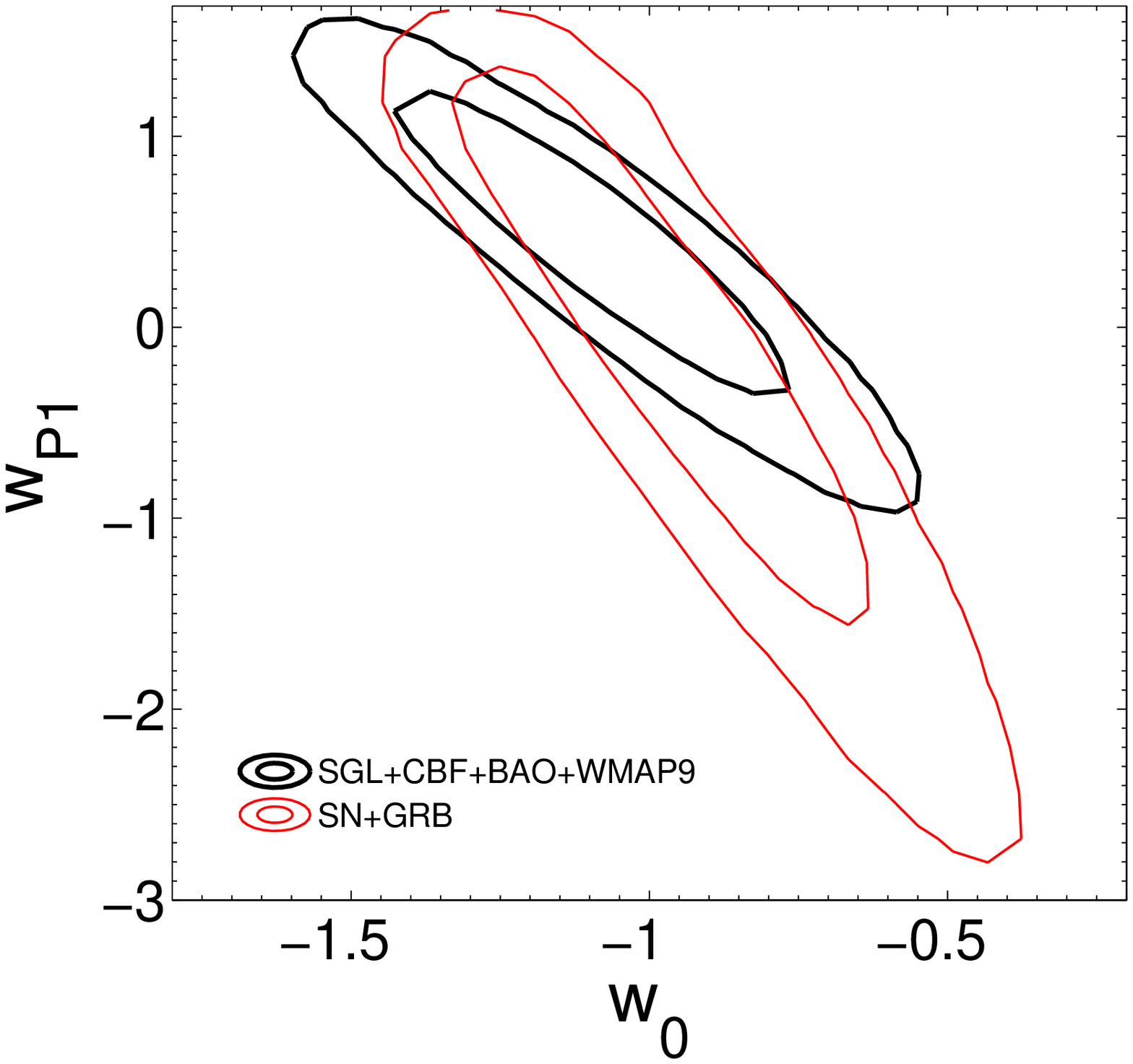}\includegraphics[angle=0,width=75mm]{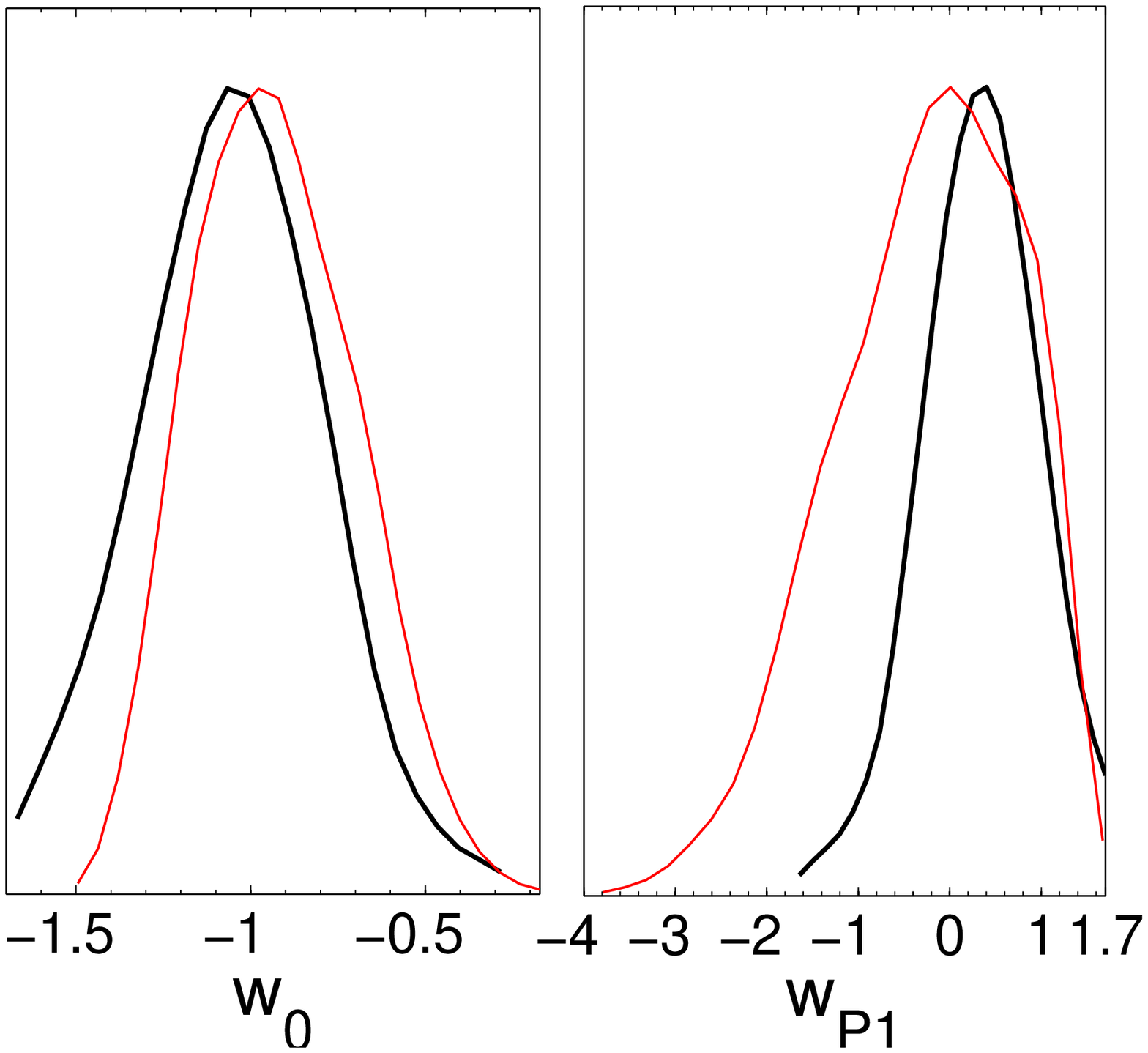}
\end{center}
\caption{ The 2-D regions and 1-D marginalized distribution with the
1$\sigma$ and 2$\sigma$ contours of parameters for CPL parametrization from the combined angular diameter distance data (Upper).
The lower panel illustrates 1$\sigma$ and 2$\sigma$ contours in the $w_0$-$w_{P1}$ parameter space and their 1-D marginalized distributions
obtained from the combined angular diameter distance data and luminosity distance data, respectively (The matter density parameters $\Omega_{b}h^2$ and $\Omega_{c}h^2$ are fixed at the WMAP9 best-fit values).
\label{CPL}}
\end{figure}

Fitting the data from the combinations of standard rulers to the CPL parametrization, the most widely explored dark energy model with variable equation of state coefficient, we get the results shown in Fig.~\ref{CPL}, where the marginalized
probability distribution of each parameter and the marginalized 2D confidence contours of parameters are presented. The best fit is
$\Omega_bh^2=0.0228\pm0.0010$, $\Omega_{c}h^2=0.1155\pm0.0055$, $w_0=-1.050\pm0.375$, $w_{P1}=0.440\pm1.250$, and $H_0=68.20\pm3.85 \rm km s^{-1} Mpc^{-1}$.
After marginalizing {\bf over} $H_0$, we obtain the dust matter density parameter $\Omega_m=0.292\pm0.035$, which is also in agreement with the earlier
comprehensive results with combined WMAP7, BAO and SN Ia analysis \cite{Komatsu11}.

In order to gain more insight into the compatibility between ADD and LD fits and make comparisons with the previous results, we choose
to place priors on the energy density parameters as Ref.~\cite{Biesiada10,Biesiada11}. In Fig.~\ref{CPL} we show the 68.3\% and 95.4\% confidence contours
in the ($w_0, w_{P1}$) plane and 1-D marginalized parameter likelihood distribution with the two dataset categories (standard ruler and
standard candle data) for $\Omega_{b}h^2=0.02264$ and $\Omega_{c}h^2=0.1138$ (the best fit of the final WMAP9 observations \cite{Hinshaw12}).
One can see that the $w$ coefficient obtained from the full ADD sample is in good agreement with the respective value derived from LD data.
One can see in Fig.~\ref{CPL} that the $\sigma$ distance between best fit values is negligibly small (i.e.
the distance among the best fits is less than 1$\sigma$). Notice that $\Lambda$CDM $(w_0,w_{P1})=(-1,0)$ is consistent with the standard candle data
and the standard ruler data at less than $1\sigma$ level. The compatibility between
fits for $w_0$ and $w_{P1}$ are greatly improved compared with the previous literature using other
independent combined analysis \cite{Lazkoz08,Biesiada10,Biesiada11}. Ref.~\cite{Biesiada10} obtained the cosmic equation of state parameters
in the CPL parametrization with a combined sample of $n=20$ strong lensing systems from Sloan Lens ACS and Lens Structure and Dynamics surveys,
and independently noticed systematic deviation between fits done on standard candles and standard rulers.
More recently, Ref.~\cite{Biesiada11} extended the analysis by combining the SGL data with the CMB acoustic peak location and BAO data,
however, differences in central values of the best-fit cosmological parameters are still visible between standard rulers and standard candles.
The \textbf{difference} in our analysis may attribute to the more precise BAO and CMB measurements, combined with other complementary astrophysical probes including SGL and CBF. For comparison we also report the values of the best-fit parameters both from the ADD and LD data in Table~\ref{result}.

\begin{figure}
\begin{center}
\includegraphics[angle=0,width=85mm]{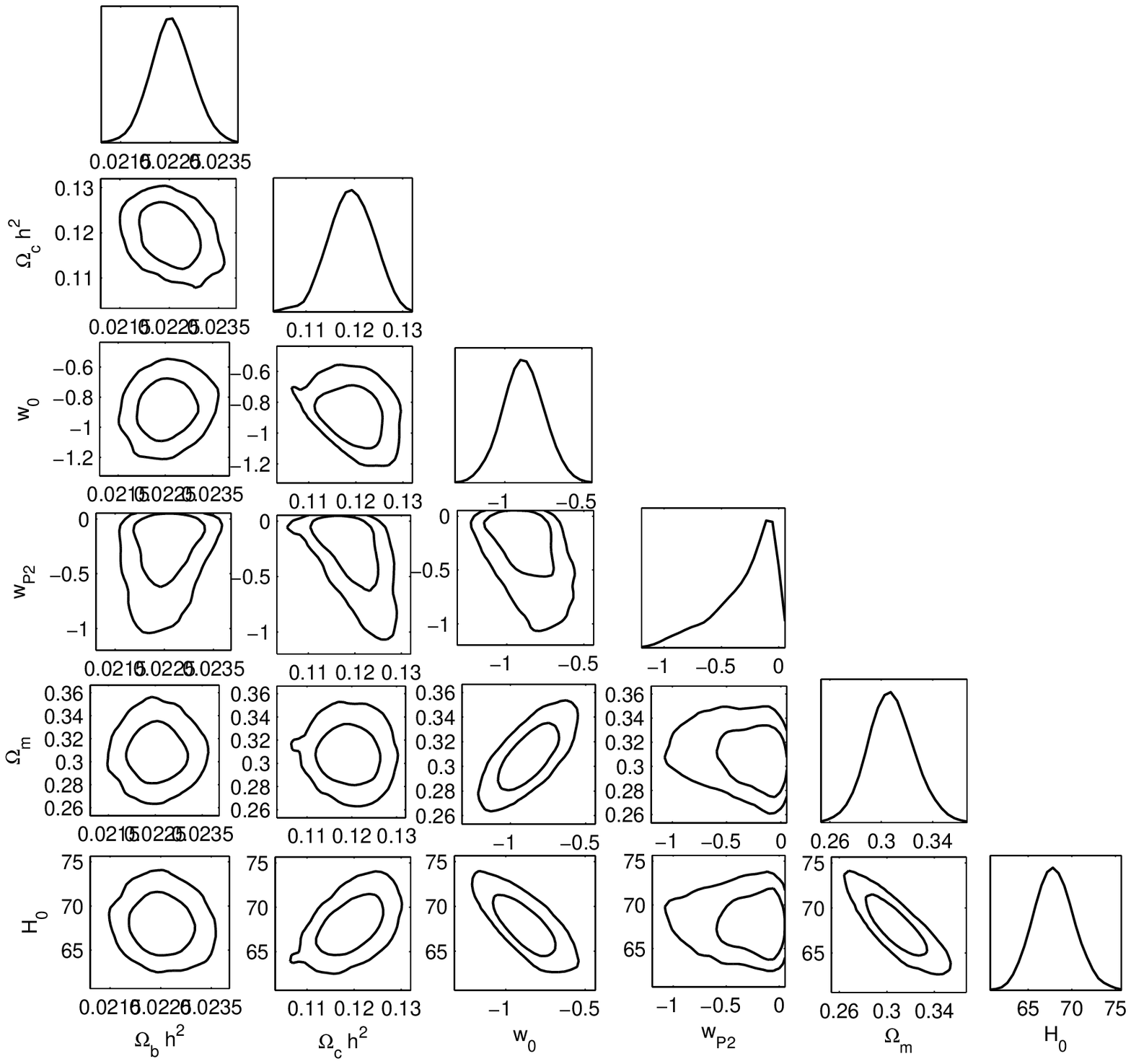}
\includegraphics[angle=0,width=75mm]{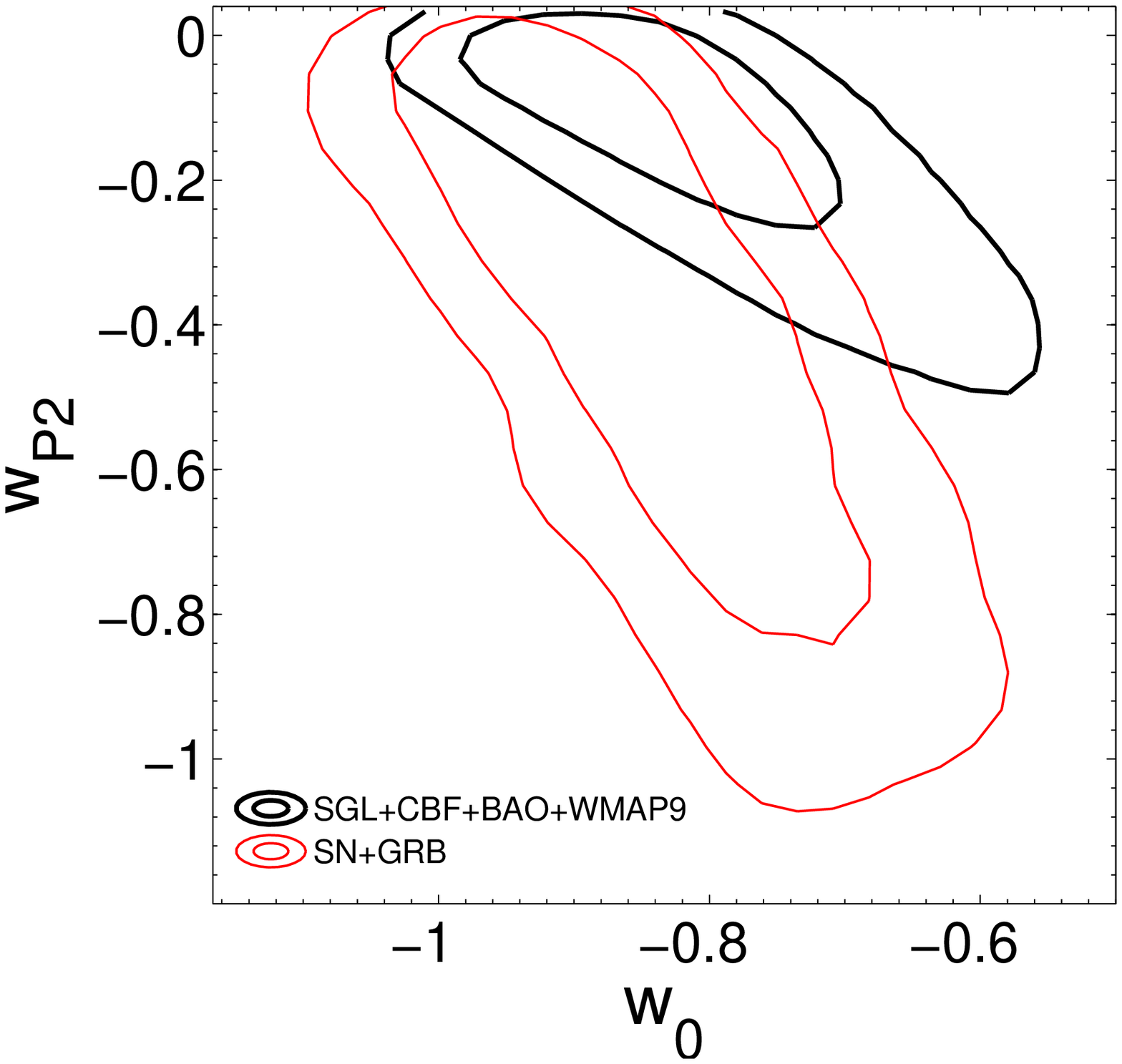}\includegraphics[angle=0,width=75mm]{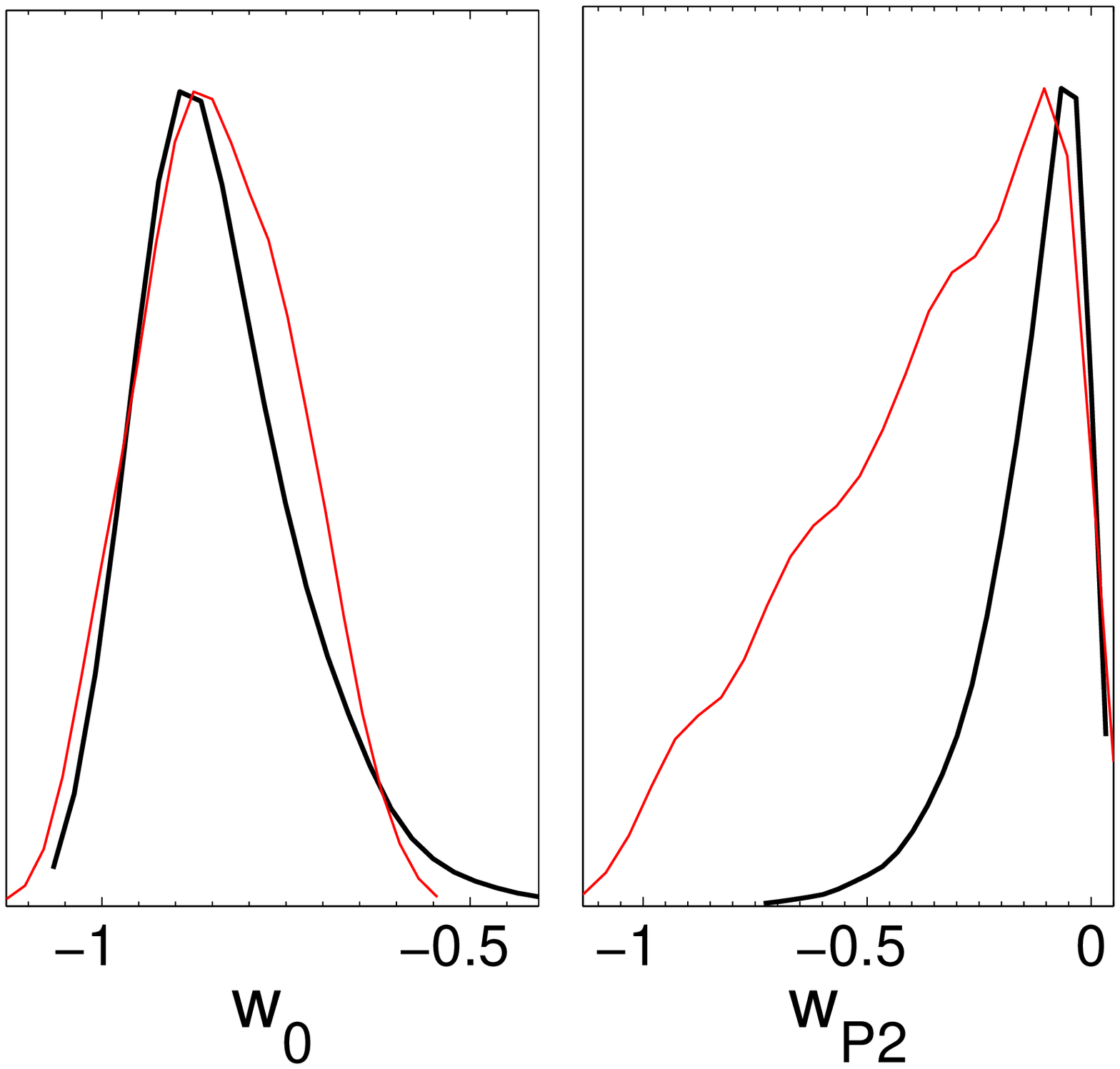}
\end{center}
\caption{ The same as Fig.~\ref{CPL}, but for the EoS parametrization $w=w_0+w_{P2}z$.
\label{model-1}}
\end{figure}

In the case of evolving equation of state in the $w=w_0+w_{P2}z$ parametrization, we obtain the results from standard
rulers as shown in Fig.~\ref{model-1}. The best fit is $\Omega_bh^2=0.0228\pm0.0011$, $\Omega_{c}h^2=0.1175\pm0.0050$ ($\Omega_m=0.305\pm0.026$),
$w_0=-0.900\pm0.200$, $w_{P2}=-0.100\pm0.325$, and $H_0=67.80\pm3.41 \rm km s^{-1} Mpc^{-1}$. Fig.~\ref{model-1} shows a comparison between the two distance data sets after assuming the WMAP9 priors on the matter fraction parameters $\Omega_bh^2$ and $\Omega_{c}h^2$. We note the $1\sigma$ confidence interval
for $w$ from the combined ADD data set lies within the $1\sigma$ CI from the LD data, which demonstrates the compatibility between the two distance
observations, although the corresponding consistency is weaker than that in the case of CPL parametrization. In this case, a possible explanation
of this tendency could be that this linear parametrization is largely redshift dependent asymptotically at high redshifts (sensitive to the ADD data especially CMB), while the LD data could play an important role in the relatively low-redshift constraints. Therefore, it is not surprising that for this EoS
parametrization, the mild difference in best fits between standard candles and standard rulers persists, as also clearly reflected by Table~\ref{result}.
Moreover, the negative central value of $w_{P2}$ fit in both joint analysis is fully compatible with $w_{P2}=0$ case when the $1\sigma$ confidence
interval is considered.

In the next subsections we will look deeper into the results by applying the two distance data to more cosmic EoS parameterizations.
In the dark energy model with generalized equation of state, the two EoS parameterizations in the above subsection could be fully recovered when
$\beta\rightarrow -1$ and $\beta\rightarrow +1$, respectively. One can deduce that these two models are different projections in different parameter
subspaces of the general EoS model. 
We choose to assign different values $\beta=\pm2$ to this parameter and obtain two time-dependent EoS parameterizations like CPL.

\subsection{Truncated GEoS model with $\beta=\pm2$}

\begin{figure}
\begin{center}
\includegraphics[angle=0,width=85mm]{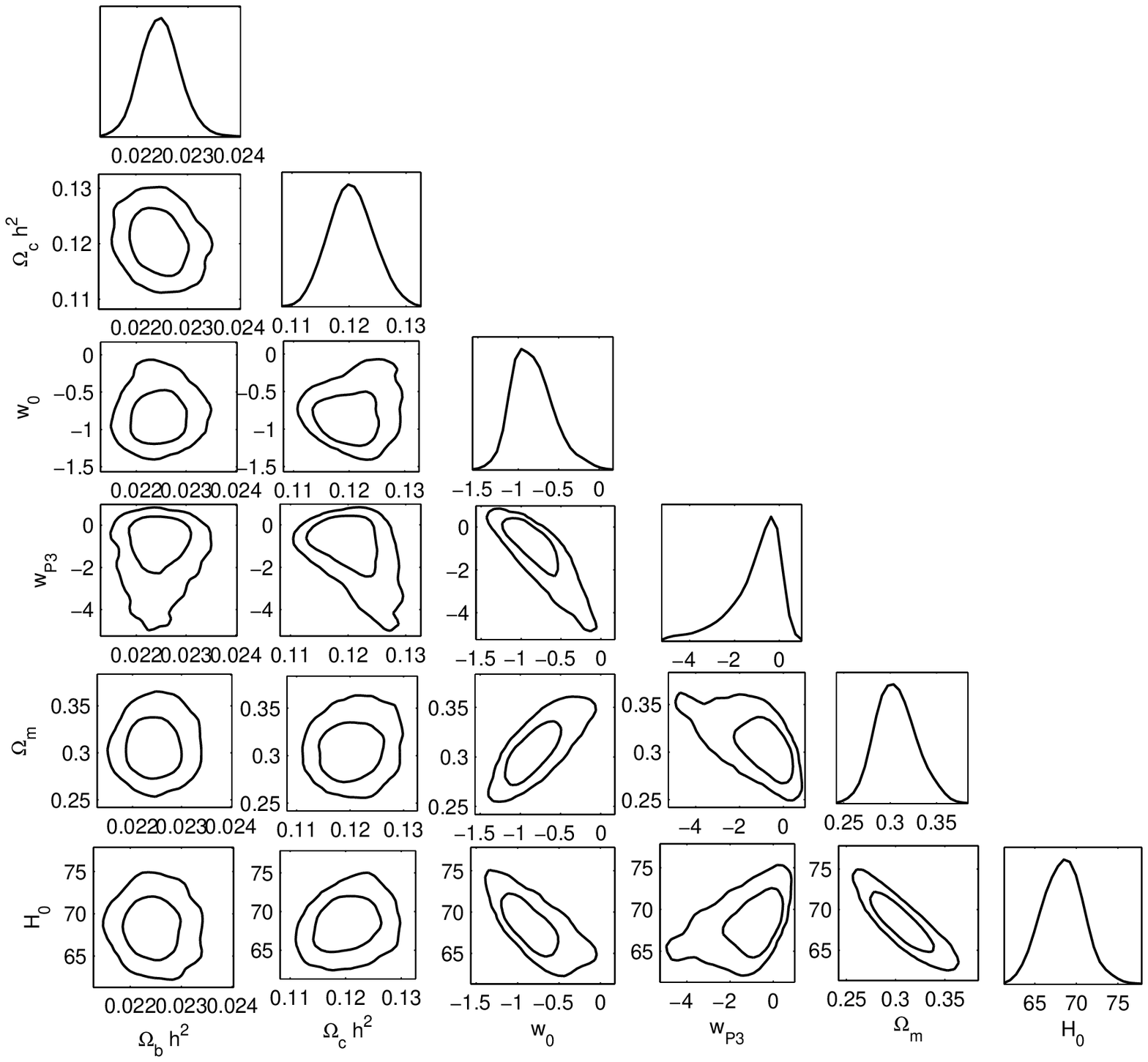}
\includegraphics[angle=0,width=75mm]{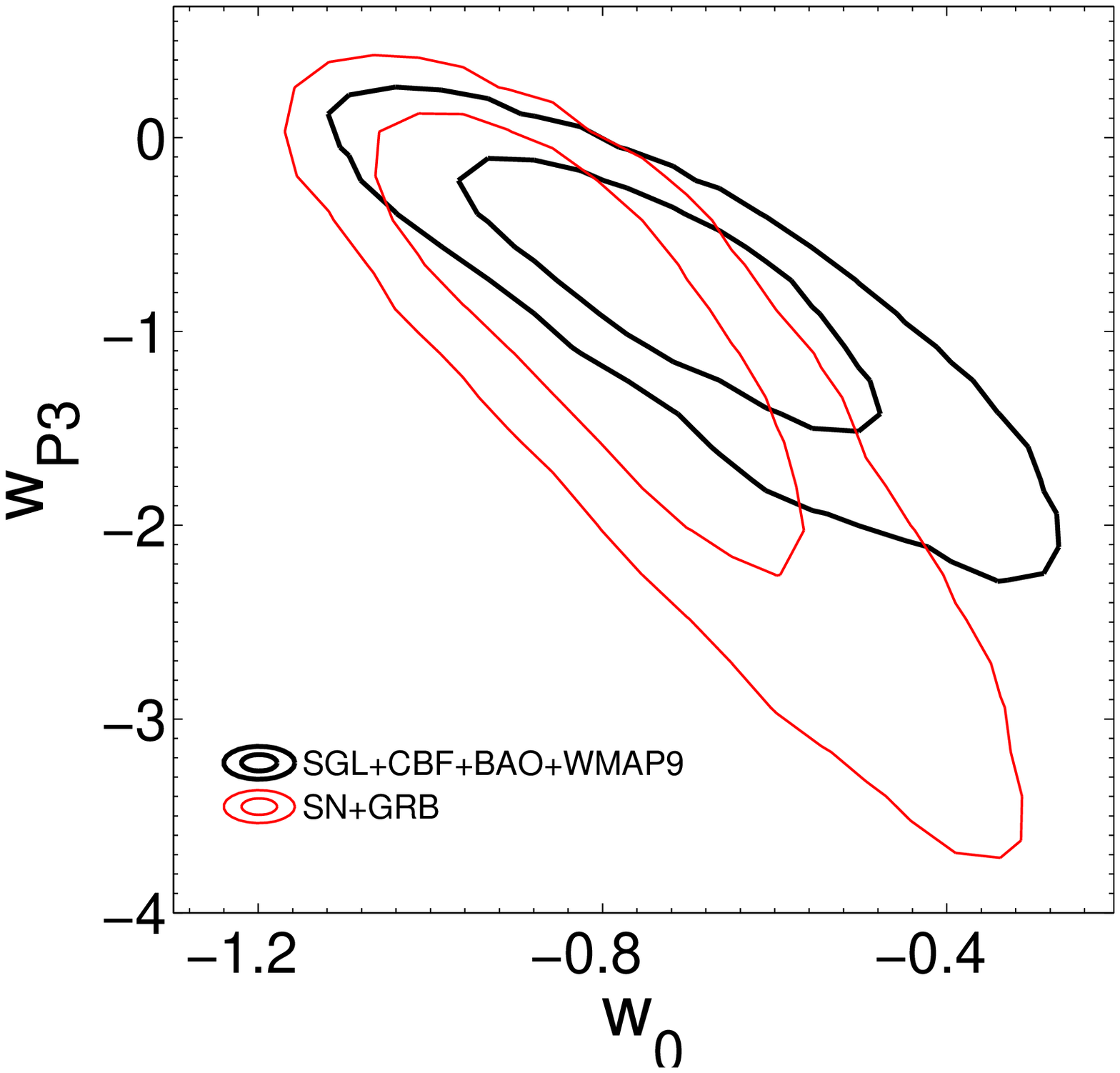}\includegraphics[angle=0,width=75mm]{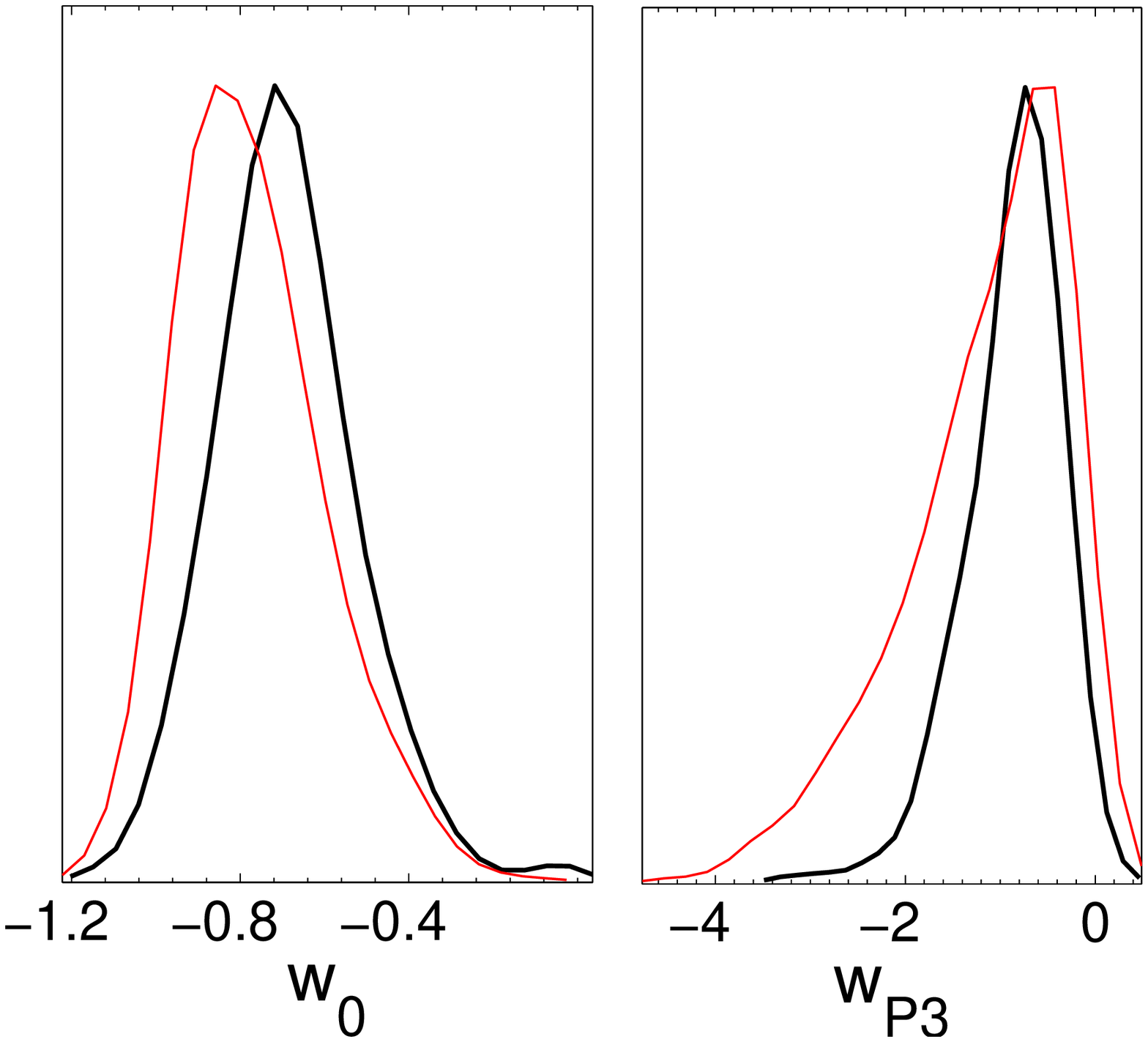}
\end{center}
\caption{The 2-D regions and 1-D marginalized distribution with the
1$\sigma$ and 2$\sigma$ contours of parameters for the truncated GEoS model with $\beta=+2$ from the combined angular diameter distance data (Upper).
Comparisons between the combined angular diameter distance data and luminosity distance data are also showed (Lower).
\label{model+2}}
\end{figure}

In the first case, we consider the $\beta=+2$ parametrization for dark energy with the following equation of state
\begin{equation}
\label{beta2}
w(z)= w_0-w_{P3}\frac{(1+z)^{-2}-1}{2}.
\end{equation}
The corresponding dimensionless dark energy density is then
\begin{eqnarray}
\label{deneq1}
\Omega_{X}(z)=(1-\Omega_{m}-\Omega_{r})\times(1+z)^{3(1+w_0+w_{P3}/2)}\exp\Big[\frac{3w_{P3}}{2}\big(\frac{(1+z)^{-2}-1}{2}\big) \Big].
\end{eqnarray}
In this model, we also have four model parameters $\textbf{p}=\{\Omega_bh^2,\Omega_{c}h^2,~w_0,~w_{P3}\}$ and a nuisance parameter $H_0$.
The results are displayed in Fig.~\ref{model+2}, with the best fit $\Omega_bh^2=0.0228\pm0.0010$, $\Omega_{c}h^2=0.1180\pm0.0040$ ($\Omega_m=0.303\pm0.033$),
$w_0=-0.965\pm0.347$, $w_{P3}=-0.340\pm1.015$, and $H_0=68.60\pm1.99 \rm km s^{-1} Mpc^{-1}$. At 68.3\% C.L., we find that this model is still compatible
with $\Lambda$CDM, i.e. the case ($w_0=-1$; $w_{P3}=0$) typically lies within outside the 1$\sigma$ boundary though very close to it.
We further use the best-fit values of $\Omega_{b}h^2$ and $\Omega_{c}h^2$ from WMAP9 to obtain the confidence contour of the EoS parameters
displayed in Fig.~\ref{model+2}. Two-dimensional and analysis performed for $w_0$ and $w_{P3}$ shows the compatibility between ADD and LD data
at 1$\sigma$ (let's notice that most part of the 1$\sigma$ confidence contours intersect).

\begin{figure}
\begin{center}
\includegraphics[angle=0,width=85mm]{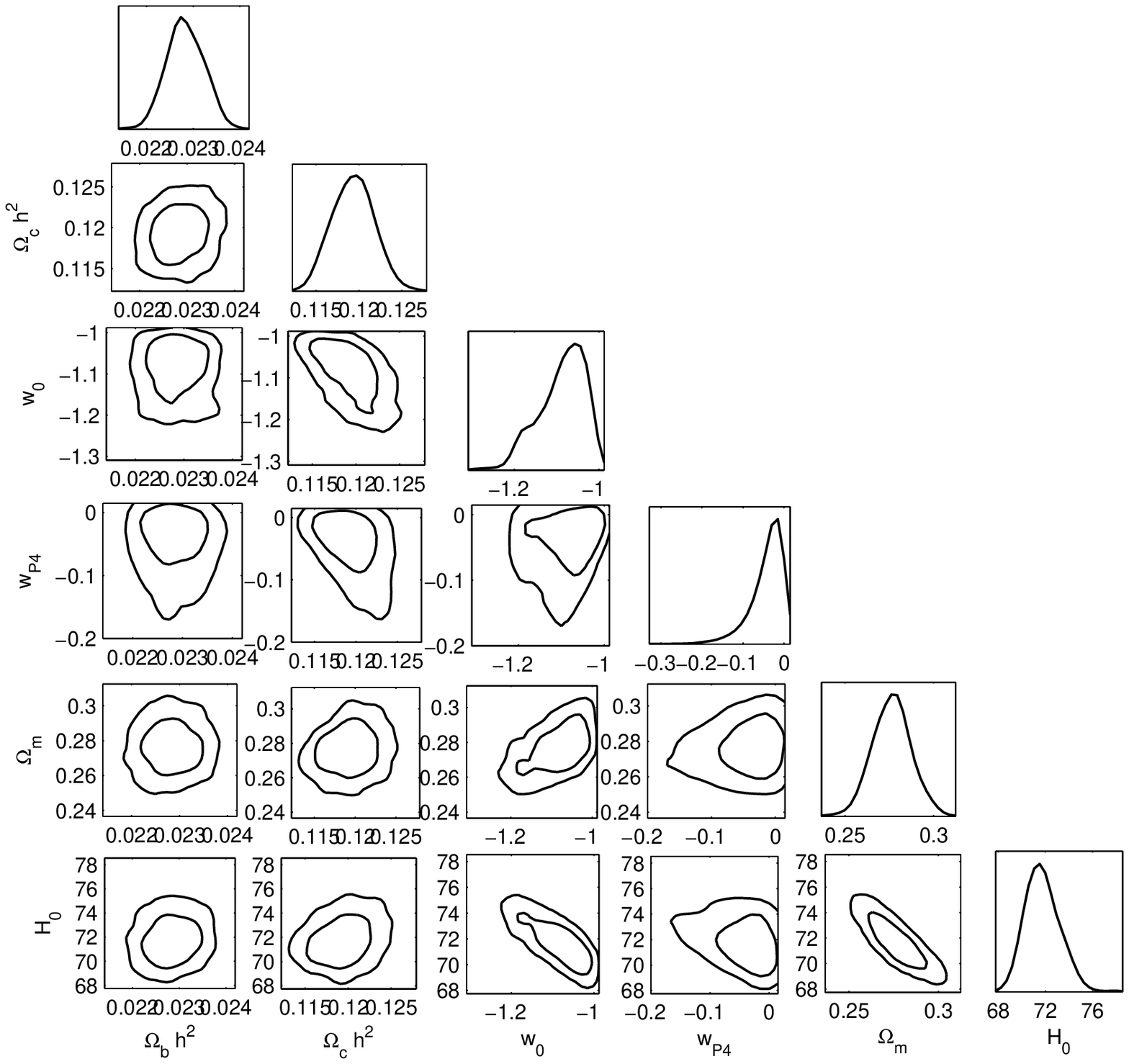}
\includegraphics[angle=0,width=75mm]{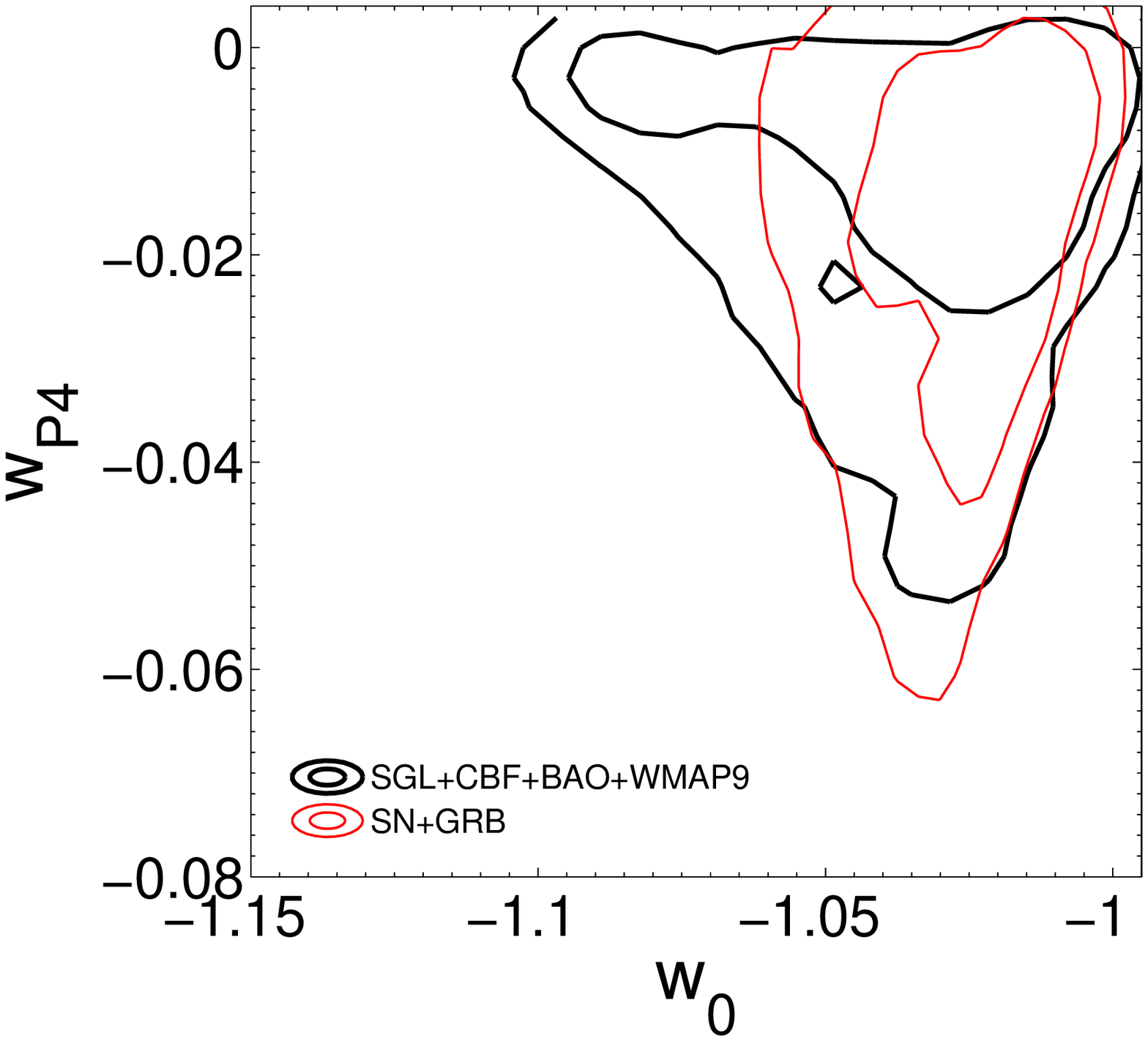}\includegraphics[angle=0,width=75mm]{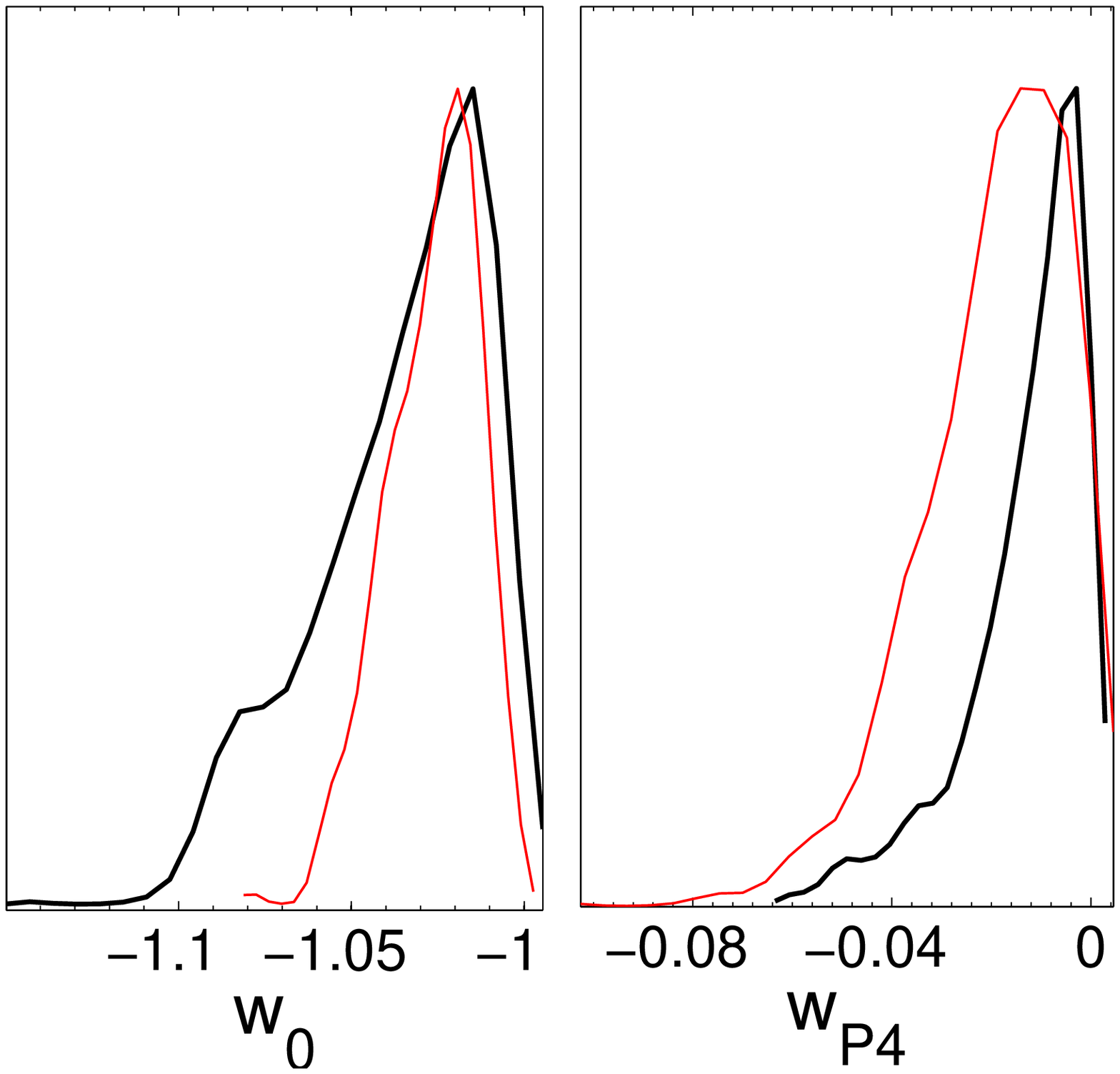}
\end{center}
\caption{The same as Fig.~\ref{model+2}, but for the truncated GEoS model with $\beta=-2$.
\label{model-2}}
\end{figure}

Performing similar analysis as before, this time with $\beta=-2$, we obtain the equation of state
\begin{equation}
\label{beta-2}
w(z)= w_0+w_{P4}\frac{(1+z)^{2}-1}{2}.
\end{equation}
and the dimensionless dark energy density
\begin{eqnarray}
\label{deneq1}
\Omega_{X}(z)=(1-\Omega_{m}-\Omega_{r})\times(1+z)^{3(1+w_0-w_{P4}/2)}\exp\Big[\frac{3w_{P4}}{2}\big(\frac{(1+z)^{2}-1}{2}\big) \Big].
\end{eqnarray}
The results are displayed in Fig.~\ref{model-2} and Table~\ref{result}. The joint analysis with standard rulers provides the best-fit parameters as
$\Omega_bh^2=0.0226\pm0.0009$, $\Omega_{c}h^2=0.1186\pm0.0036$ ($\Omega_m=0.277\pm0.018$), $w_0=-1.055^{+0.053}_{-0.133}$, $w_{P4}=-0.015^{+0.030}_{-0.075}$,
and $H_0=71.50\pm2.25 \rm km s^{-1} Mpc^{-1}$. In particular, compared with the case with $\beta=+2$, when $\beta=-2$, both of the two EoS parameters $w_0$ and $w_{P4}$ will be much more stringently constrained and the distance between ADD and LD best fits gets greatly reduced. From our analysis, the constraints of
ADD and LD data are both restrictive at the confidence level of 68.3\% and the best-fit results exhibit strong statistical agreement
between ADD and LD constraints on the cosmic equation of state, which is indicative of a strong consistency between the standard ruler
and standard candle data sets. Another evidence highlighting the equivalence between the two cosmological distance data is,
that the estimated values of $w_0$ and $w_{P4}$ within $1\sigma$ are almost identical ($1\sigma$ contours match each other perfectly).

\begin{figure}
\begin{center}
\includegraphics[angle=0,width=55mm]{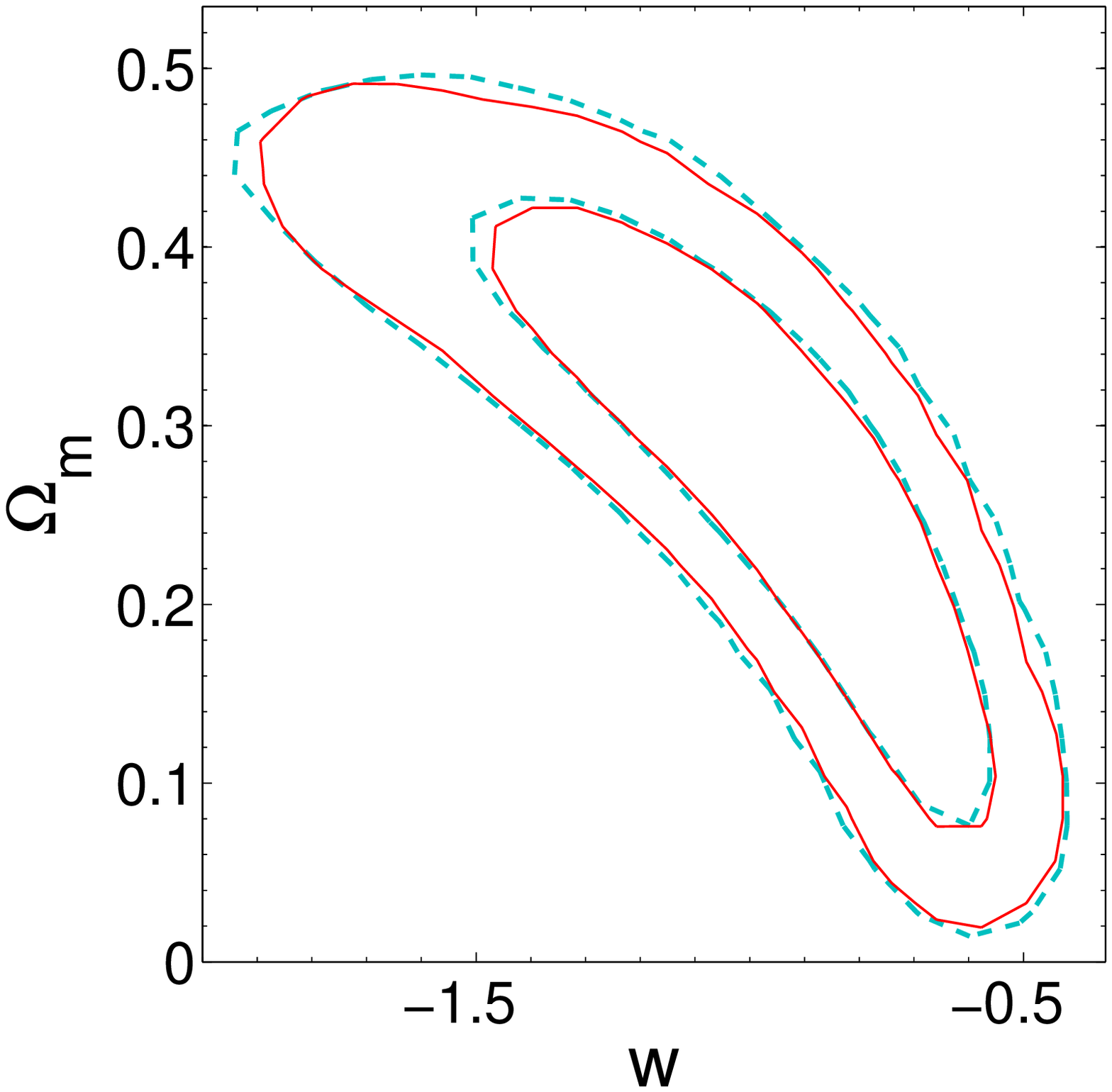}
\includegraphics[angle=0,width=55mm]{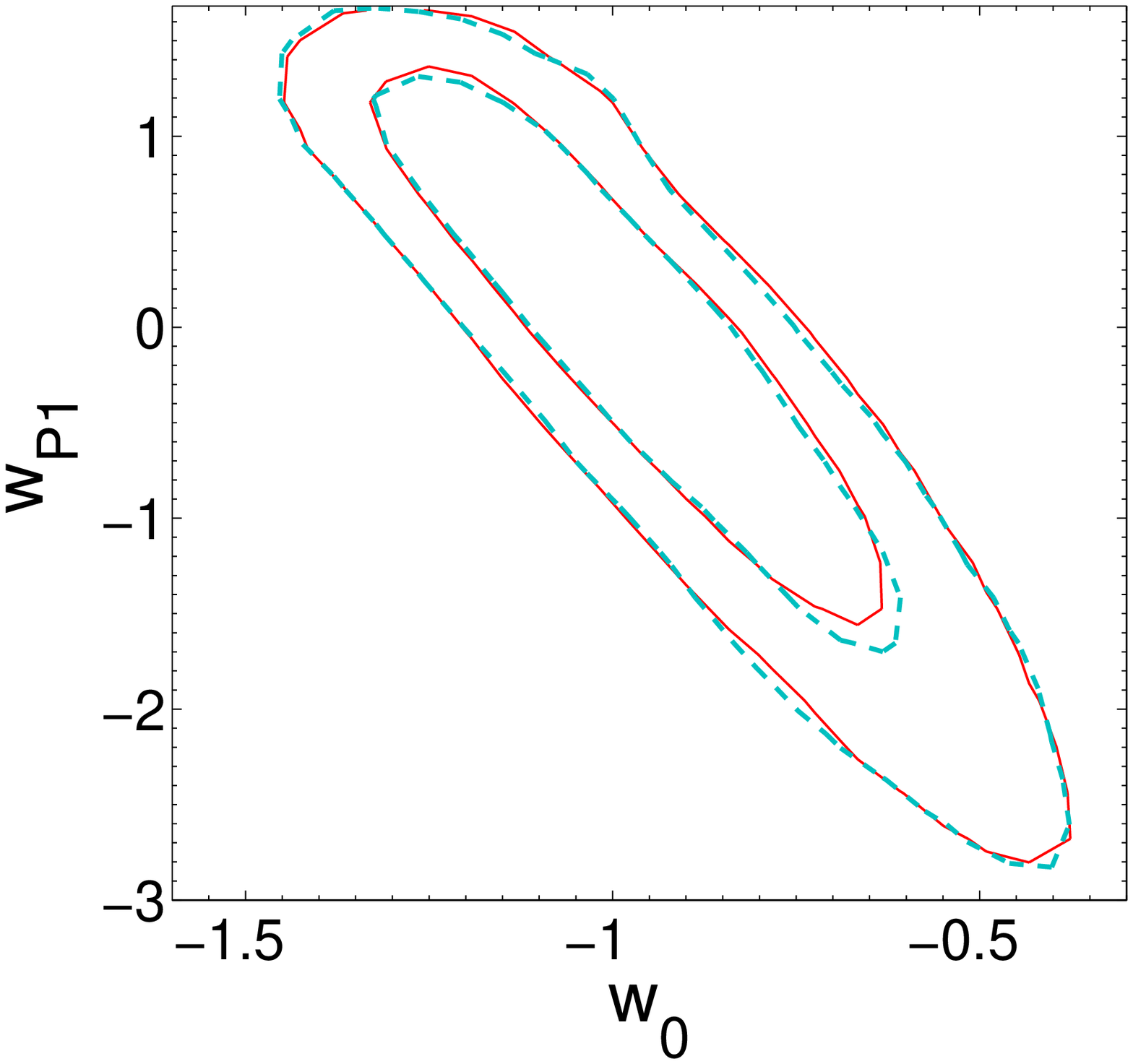}\includegraphics[angle=0,width=55mm]{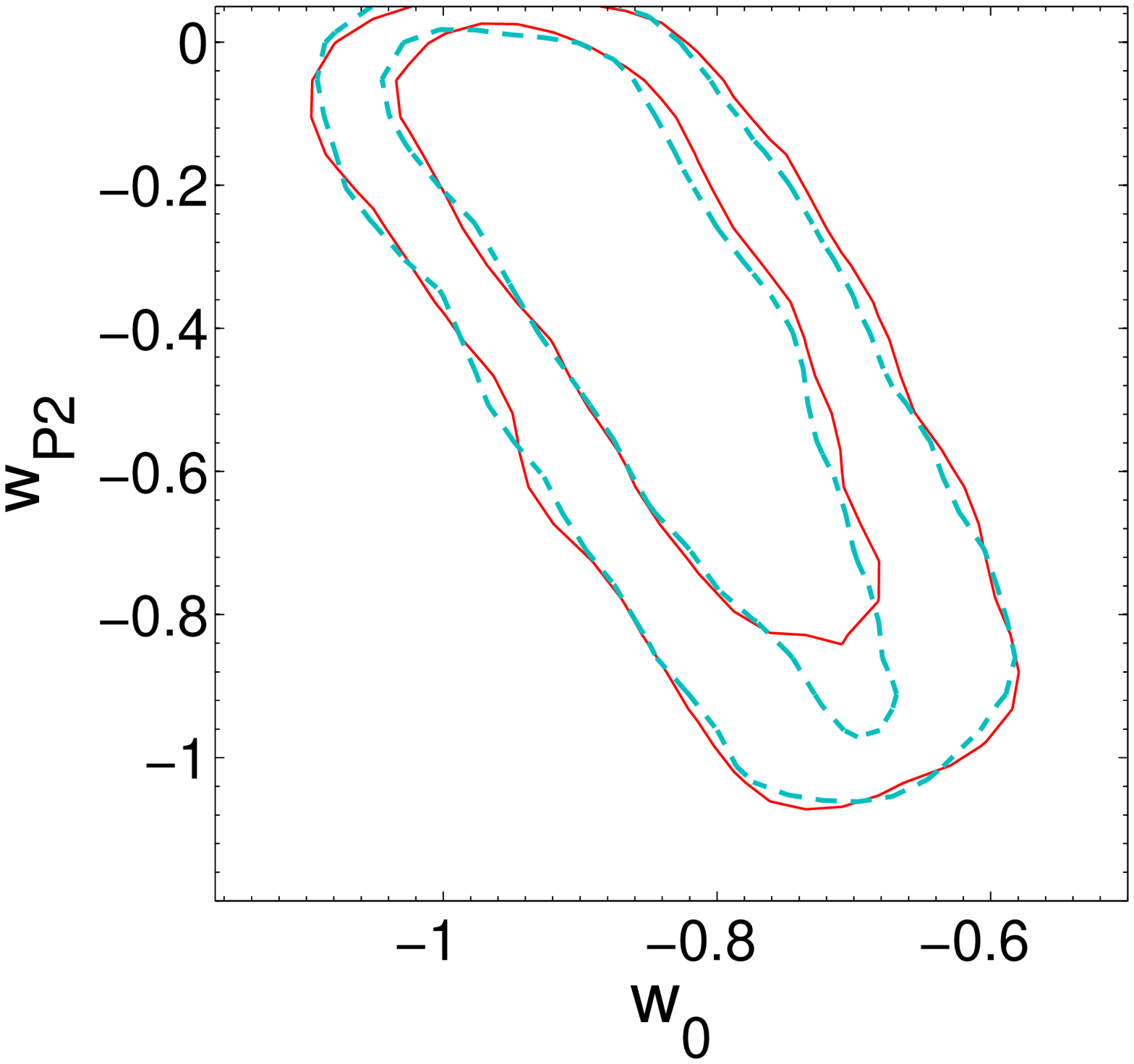}
\end{center}
\caption{ The constraint results from luminosity distance data: SN (Green line) and SN+GRB (Red line) on three
parameterizations of cosmic equation of state.
\label{SNvsGRB}}
\end{figure}

Now we choose to make two comparisons as follows. Firstly, we throw out the Gamma-Ray burst distances
and use only the SN Ia data to rederive the best estimate on the EoS parameters. Constraints on three parameterizations of cosmic equation of state
with two luminosity distance data are shown in Fig.~\ref{SNvsGRB}. Comparing the constraints from SN data with those from SN+GRB data, we find the two plots are almost the same, confirming that the current GRBs data are consistent with the SN observation, although they contribute little to the existing LD constraints.

Secondly, we use the Planck first data release \cite{Planck1} instead of WMAP9 to check its
effect on the constraints obtained in the previous section. Like WMAP9, the reduced Planck data we use is the mean values and covariance matrix of
$\{R, l_a, \Omega_b h^2\}$ from the Planck temperature data combined with Planck lensing, as well as WMAP
polarization at low multipoles ($l \leq 23$)\cite{Wang13}, which represents the tightest constraints from CMB data at present.
Note the original Planck+lensing+WP data derived from the Planck archiv data include the mean values and covariance matrix of
the data set $\{R, l_a, \Omega_b h^2, n_s\}$, where $n_s$ is the powerlaw index of primordial matter power spectrum.
The Gaussian distributions of the four reduced data points are given with the following means and standard deviations $\sigma$ \cite{Wang13}
\begin{eqnarray}
\hspace{-.5cm}\bar{\textbf{P}}_{Planck} &=& \left(\begin{array}{c}
{\bar l_a} \\
{\bar R}\\
{\bar \Omega_b h^2}\\
{\bar n_s}\end{array}
  \right)=
  \left(\begin{array}{c}
301.57\pm 0.18\\
1.7407\pm 0.0094\\
 0.02228\pm 0.00030\\
0.9662\pm0.0075
\end{array}
  \right).
 \end{eqnarray}
The normalized covariance matrix of $(l_a, R, \omega_b, n_s)$ is \cite{Wang13}
\begin{eqnarray}
\mbox{NormCov}_{Planck}=\left(
\begin{array}{cccc}
   1.0000  &    0.5250  &   -0.4235  &   -0.4475    \\
  0.5250  &     1.0000  &   -0.6925  &   -0.8240    \\
 -0.4235  &   -0.6925  &     1.0000  &    0.6109    \\
 -0.4475 &   -0.8240  &    0.6109 &     1.0000  \\
\end{array}
\right)
\label{eq:normcov_planck}
\end{eqnarray}
In order to obtain the covariance matrix for $(l_a, R, \omega_b)$, we choose to marginalize the CMB distance priors
over $n_s$ as
\begin{eqnarray}
\mbox{C}_{Planck}(p_i,p_j)=\sigma(p_i)\, \sigma(p_j) \,\mbox{NormCov}_{Planck}(p_i,p_j),
\label{eq:Planckcov}
\end{eqnarray}
where $i,j=1,2,3$. The contribution of the Planck data to the $\chi^2$ value is
\begin{eqnarray}
\chi^2_{Planck}=\Delta
\textbf{P}_{Planck}^\mathrm{T}\textbf{C}_{Planck}^{-1}\Delta\textbf{P}_{Planck},
\end{eqnarray}

\begin{figure}
\begin{center}
\includegraphics[angle=0,width=50mm]{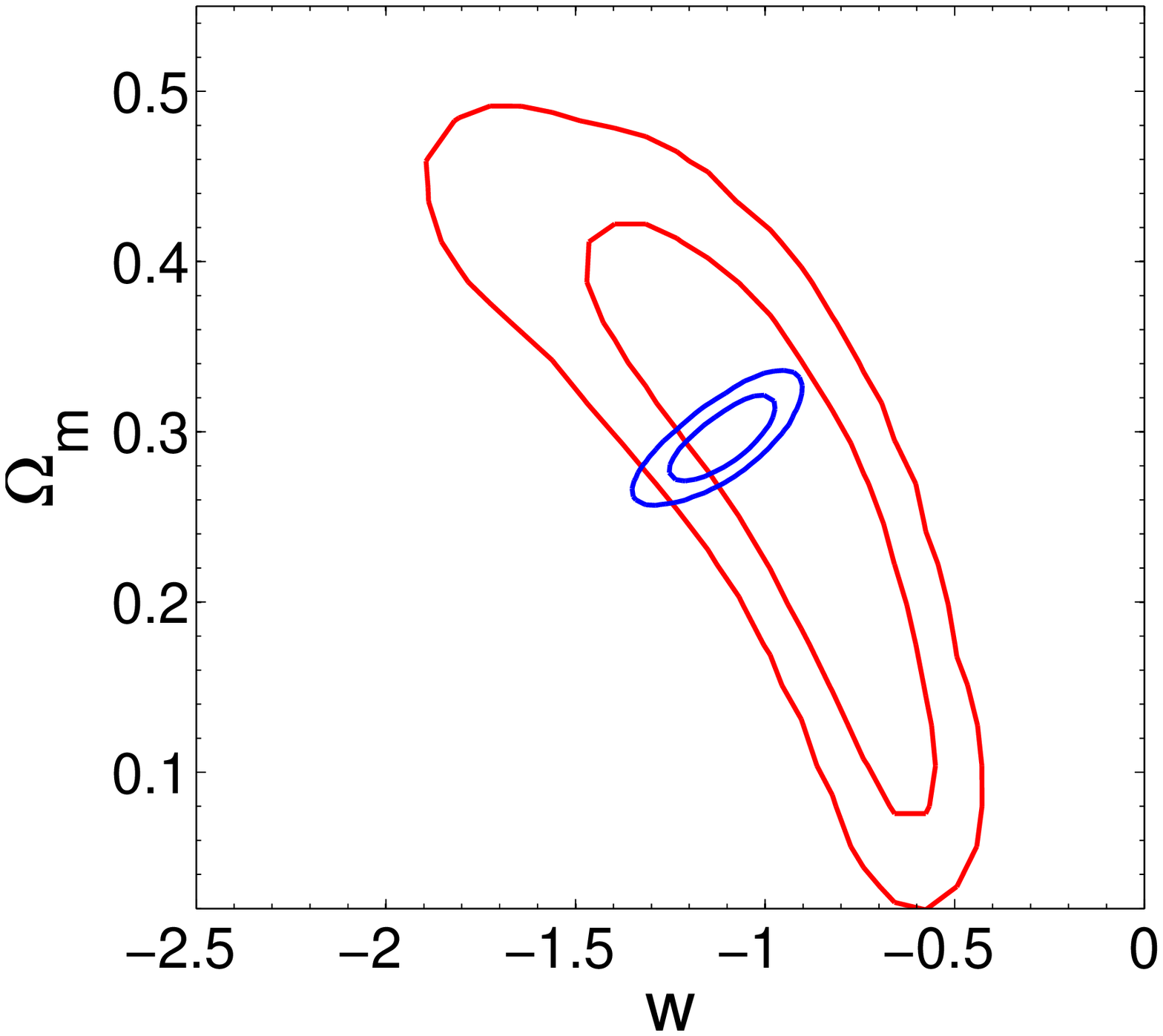}\includegraphics[angle=0,width=50mm]{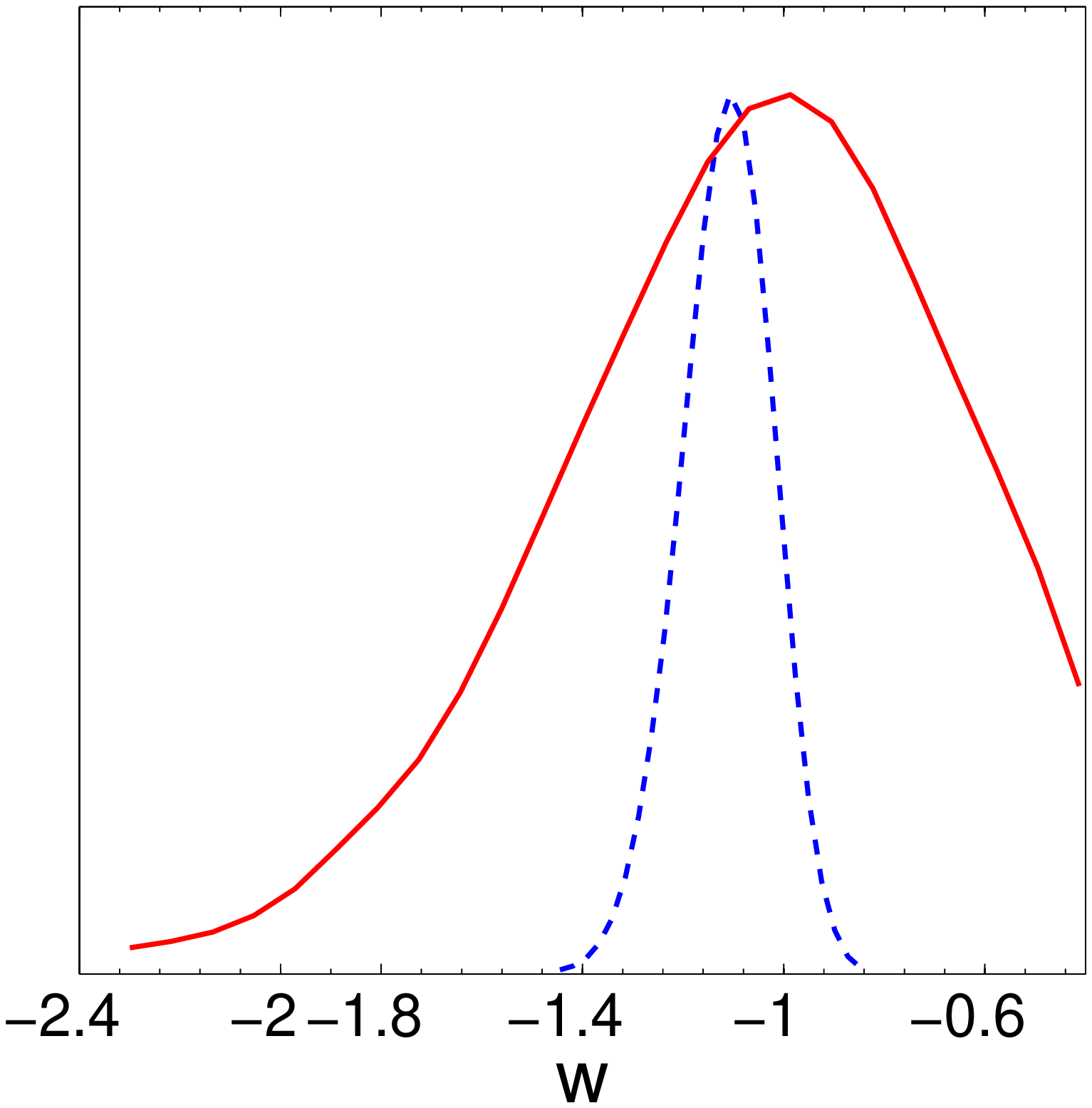}
\includegraphics[angle=0,width=50mm]{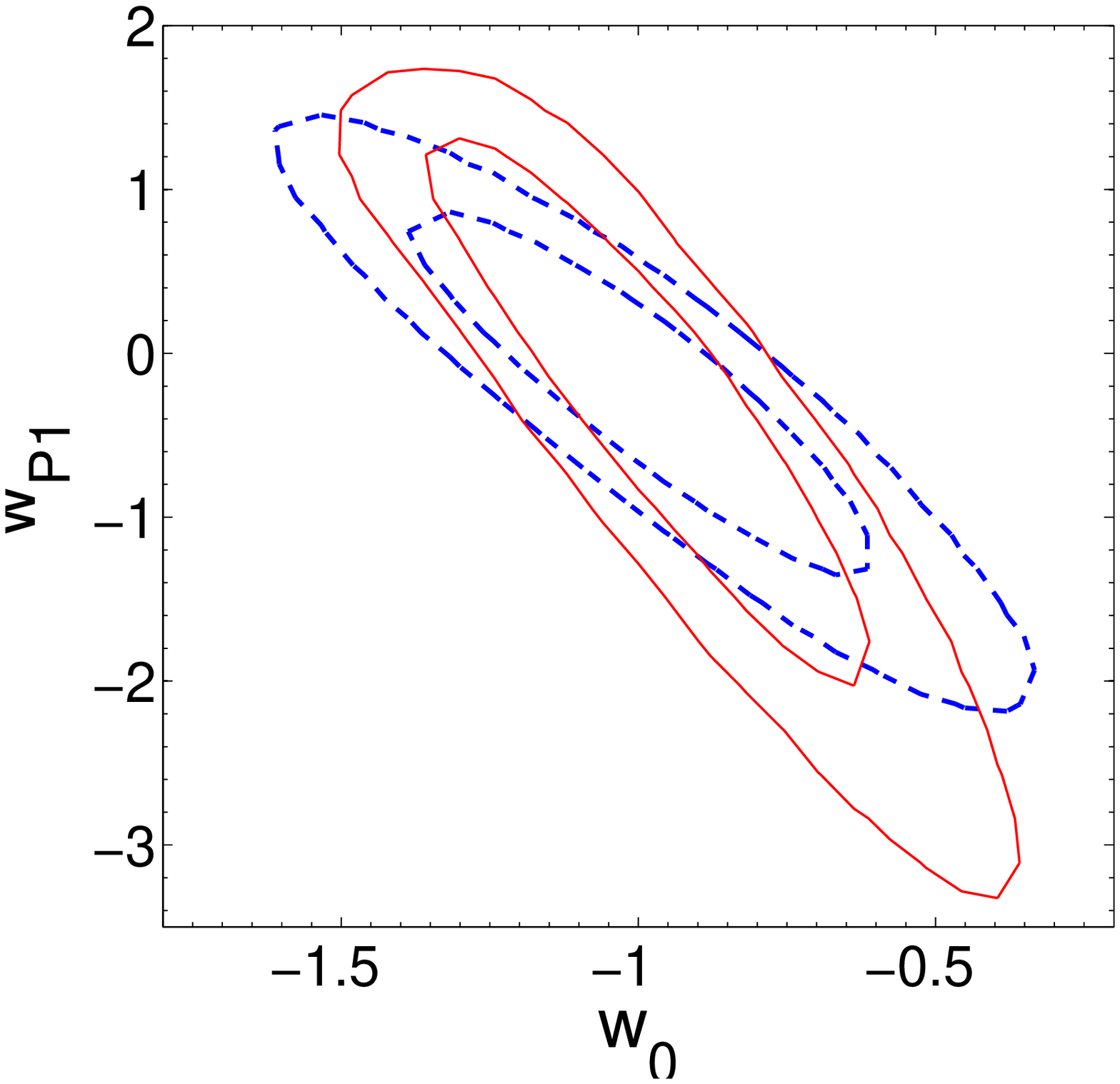}\includegraphics[angle=0,width=50mm]{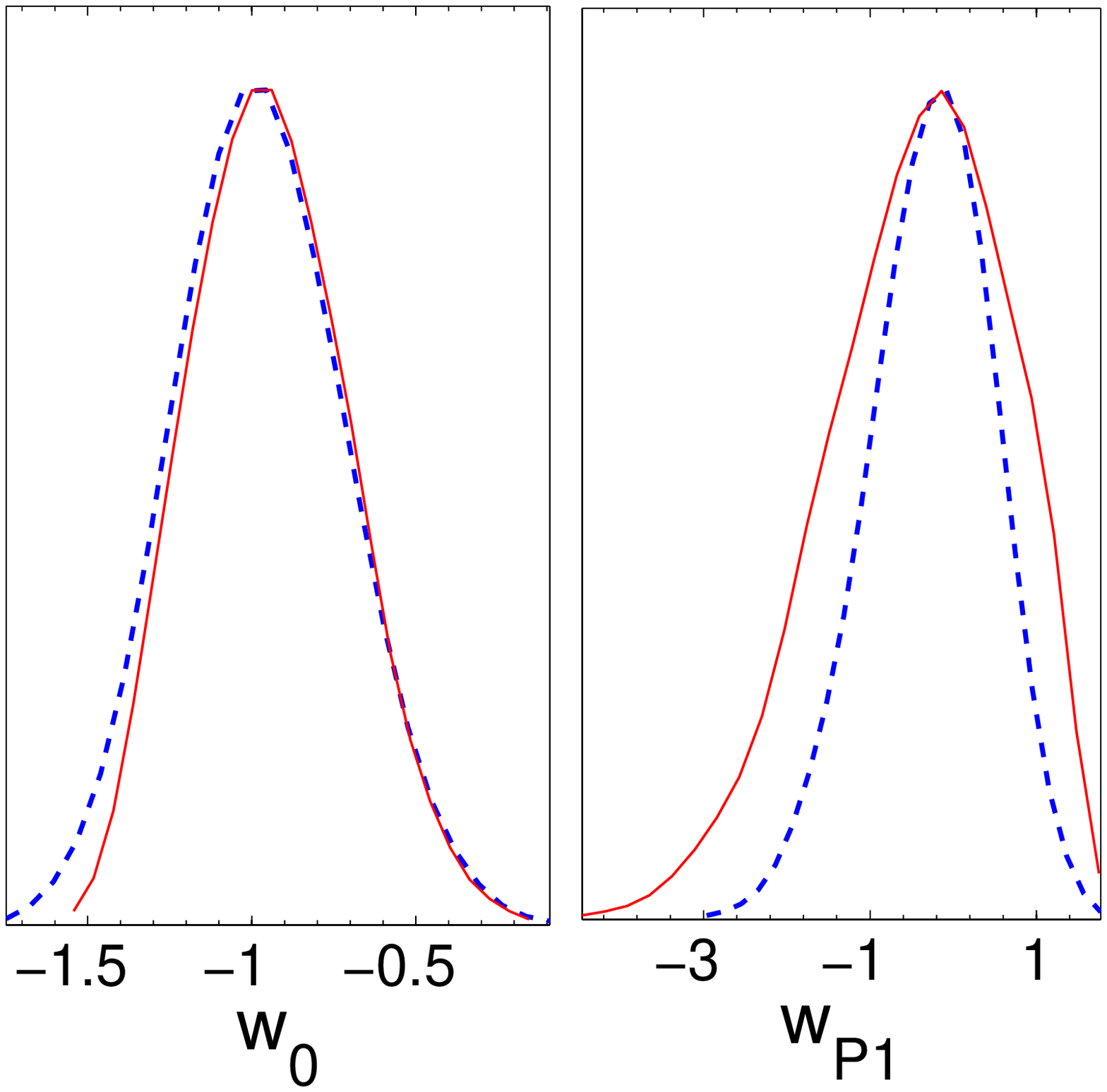}
\includegraphics[angle=0,width=50mm]{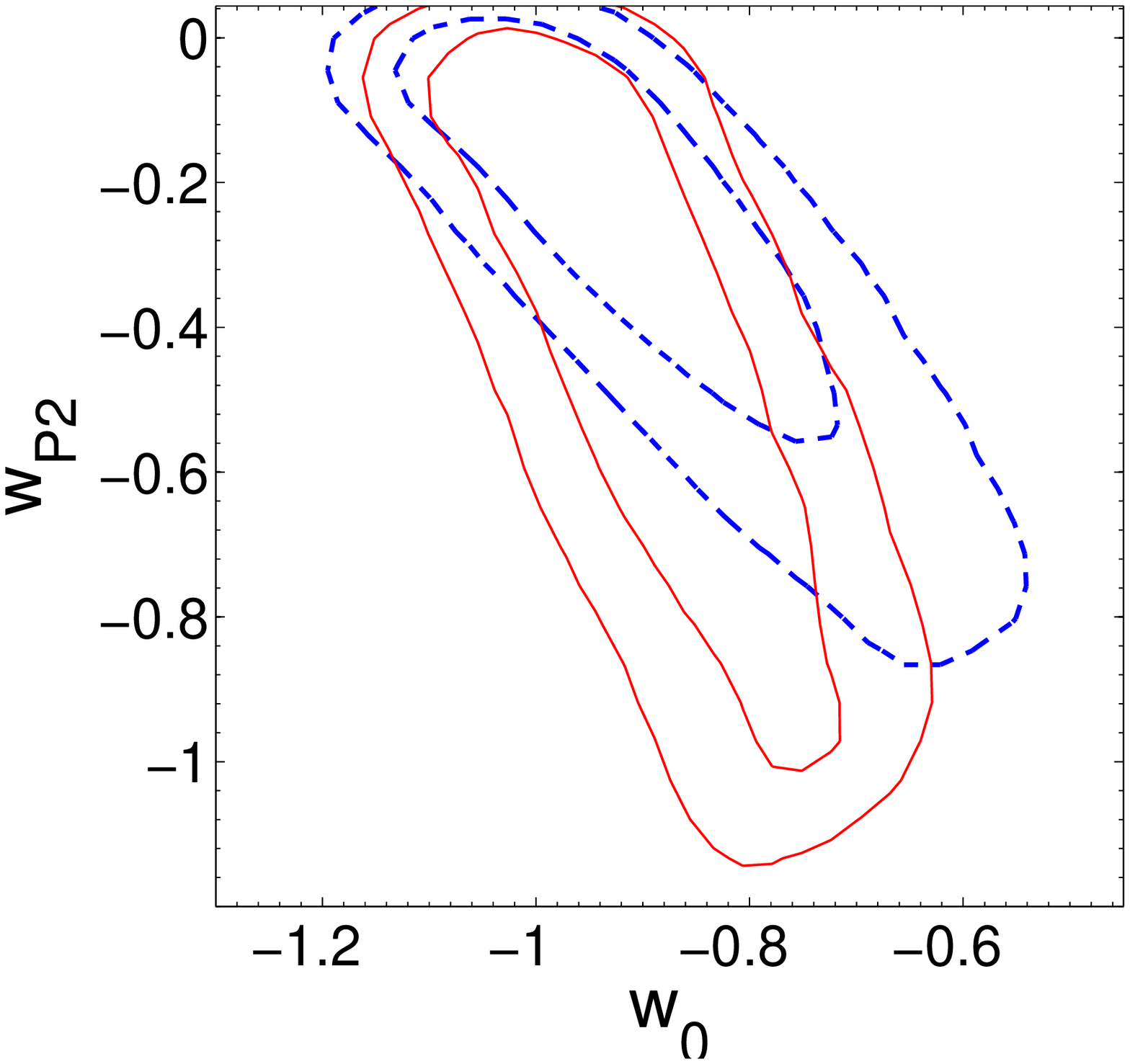}\includegraphics[angle=0,width=50mm]{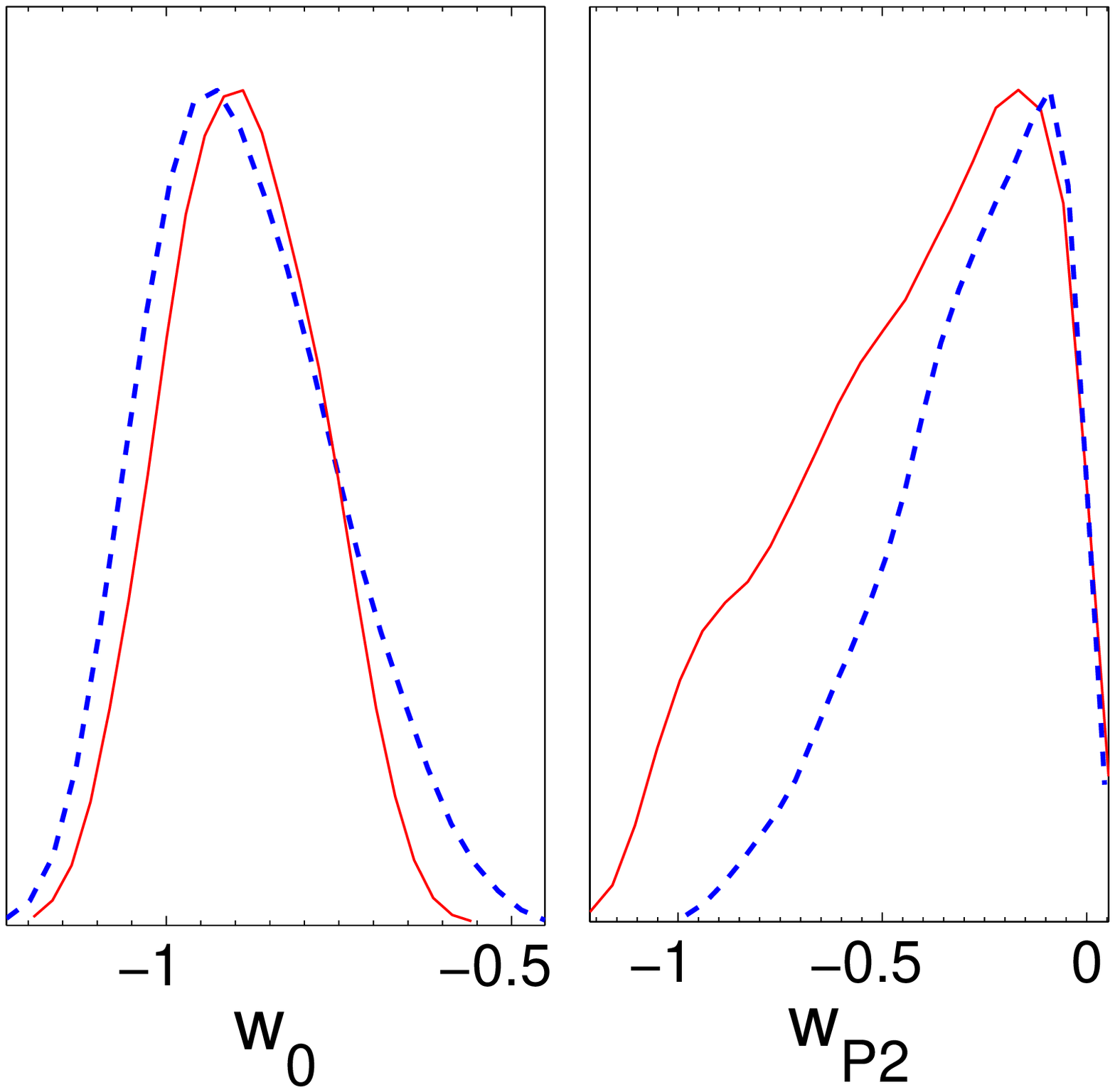}
\includegraphics[angle=0,width=50mm]{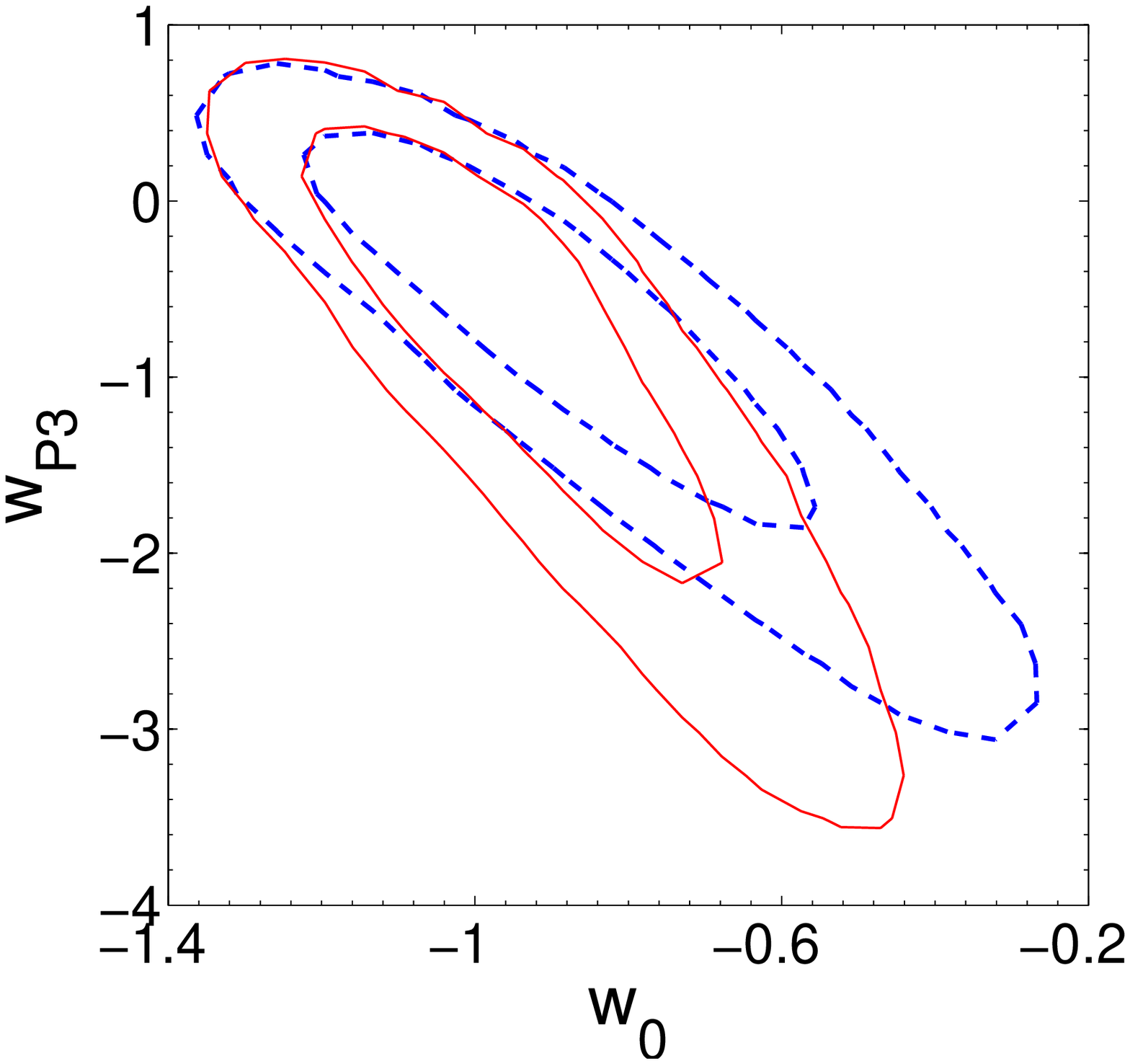}\includegraphics[angle=0,width=50mm]{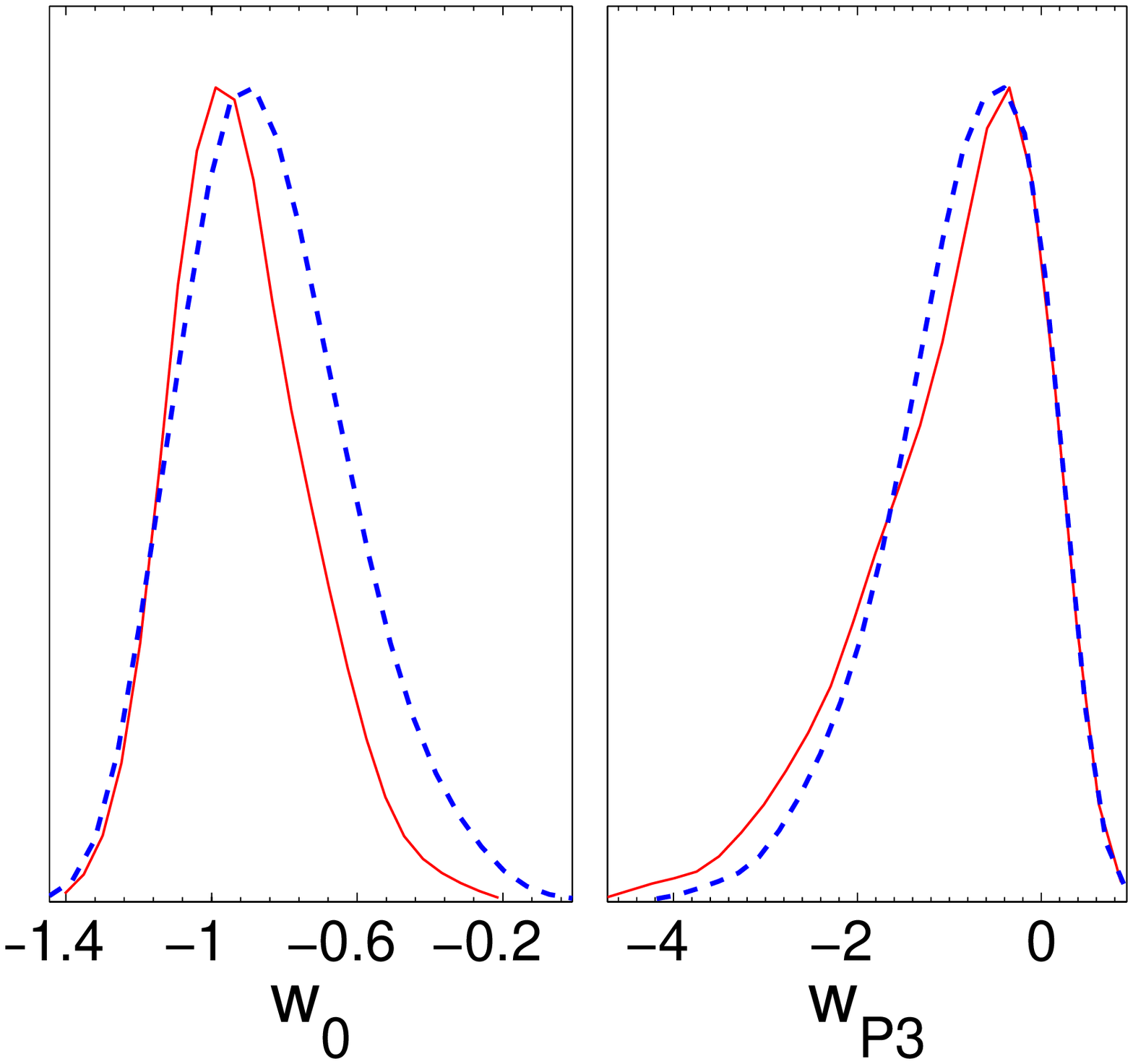}
\includegraphics[angle=0,width=50mm]{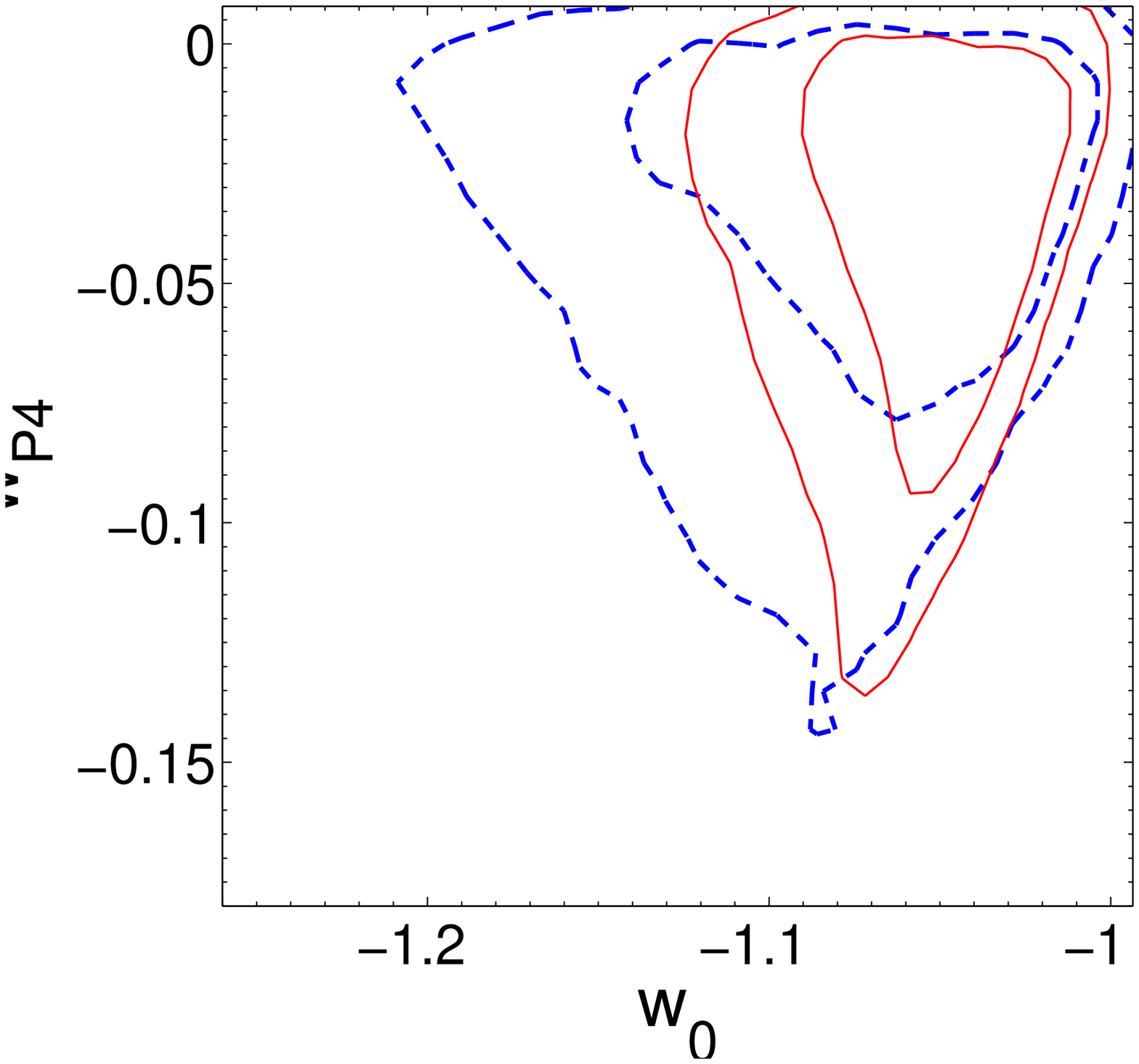}\includegraphics[angle=0,width=50mm]{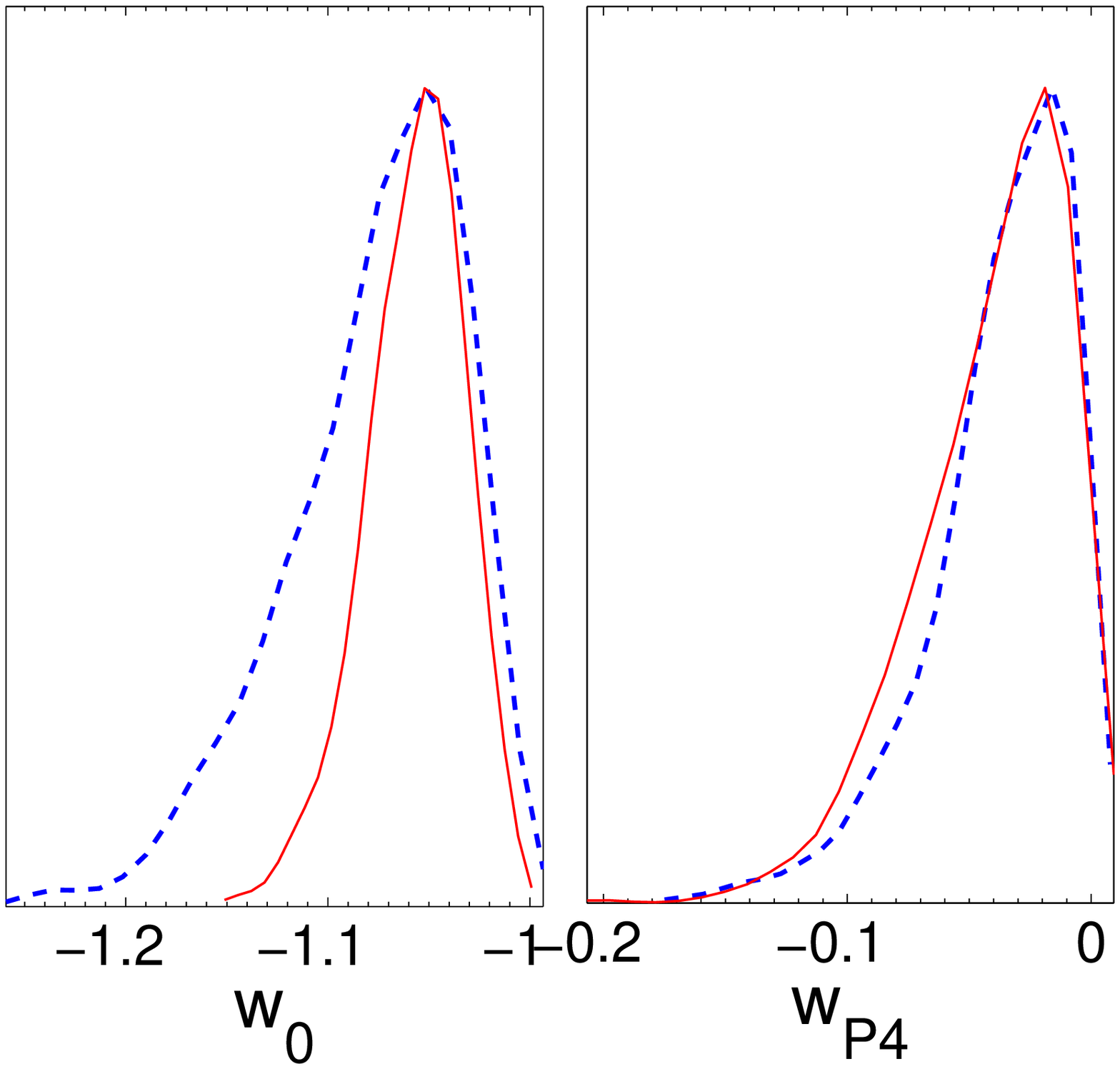}

\end{center}
\caption{The constraint results from the combined angular diameter distance data (SGL+CBF+BAO+Planck)(Blue line) and luminosity distance data (SN+GRB) (Red line). The matter density parameters $\Omega_{b}h^2$ and $\Omega_{c}h^2$ are fixed at the Planck best-fit values.
\label{Planck}}
\end{figure}

We perform analysis with SGL+CBF+BAO+Planck likelihood combinations and the results are shown in Table~\ref{result}.
The high value of $\Omega_m$ is consistent with the parameter analysis described by the Planck analysis \cite{Planck1}.
For comparison, the constraint results from the combined angular diameter distance data (SGL+CBF+BAO+Planck) and luminosity distance data (SN+GRB) are also shown in Fig~\ref{Planck}. We find that Planck data give very similar results as WMAP9 data on the EoS parameter $w(z)$, and adding Planck priors to ADD data leads to a more evident consistency with the LD data for the five EoS parameterizations.
The best-fit values of the parameters along with their $1\sigma$ uncertainties from the two different distance data are also listed in Table~\ref{result}.

Now, from the above comparison and previous works, we would like to comment on the constraint compatibility between ADD data
from standard rulers and LD data from standard candles. In our analysis with larger ADD data sets, we obtain the cosmic equation of state parameters {\bf whose values} generally agree with LD results already
known in the literature. Our results from the observational ADD data are reliable and the results are consistent with those using the LD data
of SN+GRB. Obviously, for some cosmic EoS parameterizations such as $w=w_0+w_{P2}z$, small systematic deviation between fits done on standard candles
and standard rulers still exits; however, our findings reveal that, for all cases the ADD and LD  $1\sigma$ overlap significantly.

\section{Dark energy with generalized equation of state} \label{sec:GEoS}

As a final check, we will investigate observational bounds on the parametric spaces $w_0-w_{\beta}-\beta$ from a statistical analysis involving four
classes of angular diameter distance data. These larger samples covering a wider range of redshift $z$ will allow us to
draw more information about the evolutionary properties of the cosmic equation of state.

\begin{figure}
\begin{center}
\includegraphics[angle=0,width=90mm]{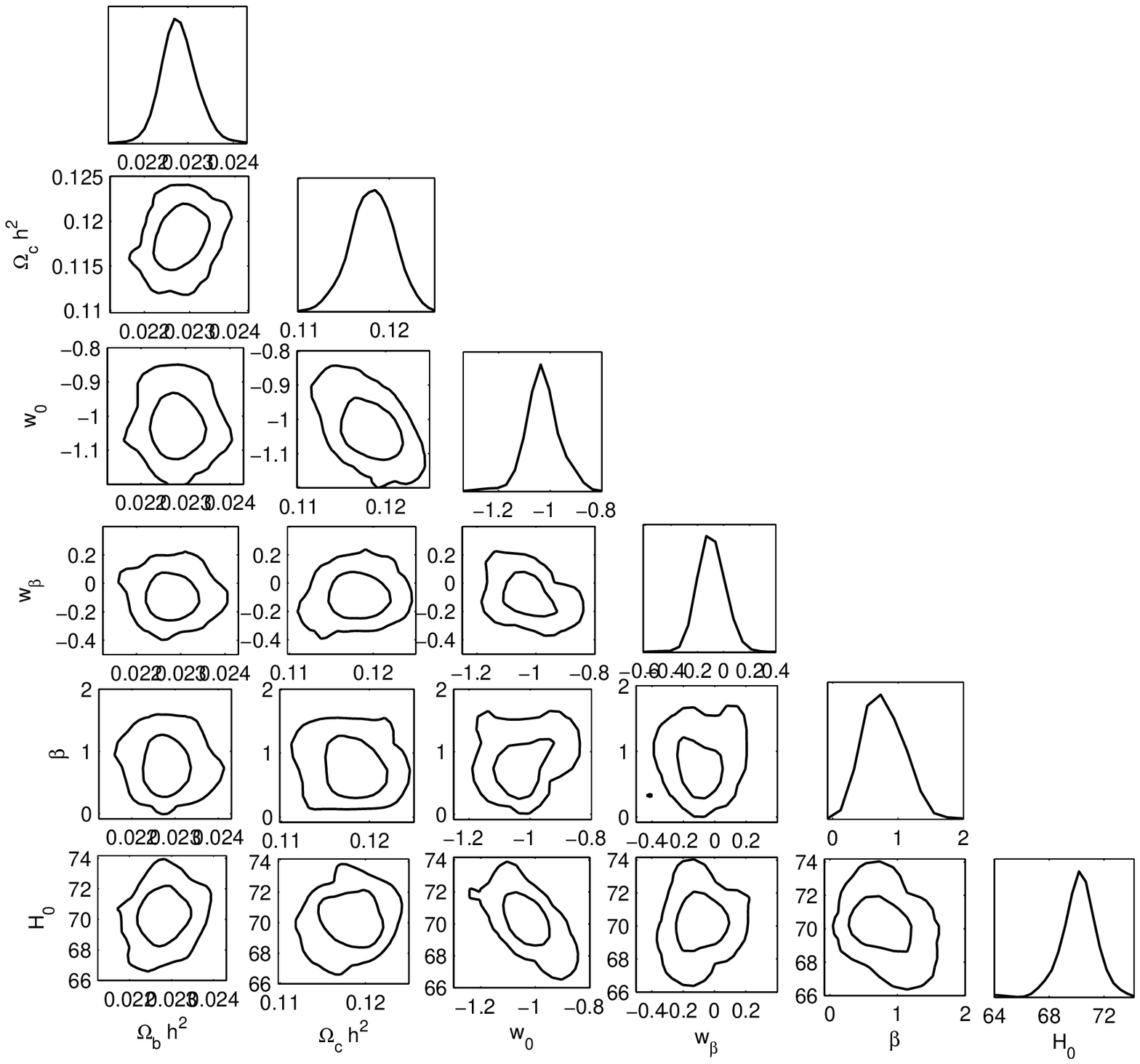}
\end{center}
\caption{ 1$\sigma$ and 2$\sigma$ confidence level contours for the GEoS parametrization from the combined angular diameter distance data (SGL+CBF+BAO+WMAP9). 1-D marginalized parameter likelihood distributions are also added.
\label{GEOS}}
\end{figure}

With the combined SGL+CBF+BAO+WMAP9 data, Fig.~\ref{GEOS} shows the marginalized probability distribution of each parameter and the marginalized 2D contour plot for the $\chi_{ADD}^2$ given by Eq.~(\ref{chiADD}). The best-fit values of the main parameters are: $\Omega_bh^2=0.0228\pm0.0007$, $\Omega_{c}h^2=0.118\pm0.004$, $\Omega_m=0.286\pm0.017$, $w_0=-1.035^{+0.110}_{-0.079}$, $w_{\beta}=-0.120^{+0.205}_{-0.115}$, and $\beta=0.751^{+0.465}_{-0.480}$, which indicates that at early times the dark energy is a subdominant component. Apparently, compared with the previous works, the combined ADD data could provide more stringent constraints on the EoS parameters $w_0$ and $w_{\beta}$ and present clear evidence supporting the CPL parametrization for the cosmic equation of state with $\beta\simeq 1$. This result marginally disagrees with the recent analysis with SN+BAO+$H(z)$: $w_0=-0.98$, $w_{\beta}=0.1$,
$\beta=-3.04$ and SN+BAO+CMB: $w_0=-1.0$, $w_{\beta}=0.28$, $\beta=0.1$ \cite{Barboza09}.

\section{Conclusions} \label{sec:conclusion}

Recent observations have provided a lot of information concerning distance measurements which is useful to analyze the dynamical behavior of the universe. However, previous studies by the others raised the question whether there is a tension between the angular diameter distance and luminosity distance data when applied to cosmography. Indeed even a very interesting remark has been made \cite{Bassett04a,Bassett04b,Uzan04} which initiated a lot of subsequent studies \cite{Cao11a,Cao11b}. Namely, even though ADD and LD are based on different concepts and give different values in physical units, they are connected with each other by the so called Etherington duality principle. The breakdown of this principle would mean that either the gravity is not a metric theory (which is so improbable that almost impossible) or there is some mechanism of non-conservation of the number of photons on the path form the source to observer (this could be as obvious as extinction or as exotic as e.g converting photons into axions) \cite{Bassett04b}. Our testing the consistency between ADD and LD can be also perceived from this perspective.

We have collected a relatively complete observational data concerning four angular diameter distance measurements to provide constraints on the cosmic equation of state, and compared the fitting results with those obtained from larger luminosity distance data. In addition to the previous probes, we use the X-ray gas mass fraction of galaxy clusters and strongly gravitationally lensed systems from various large systematic gravitational lens surveys and galaxy clusters.
As an extension of the previous works, the newly updated high redshift Gamma-Ray Bursts (GRBs) dataset is also included as complementary to SN Ia in the role of standard candles.

We have performed joint analysis of three classes of cosmological models invoked to explain accelerating expansion of the universe:
(1) Constant dark energy equation of state $w(z)=w$; (2) Variable dark energy equation of state, parametrized by $w(z) = w_0 + w_{P1}z/(1+z)$ and $w(z)= w_0+w_{P2}z$;
(3) Generalized dark energy equation of state $w(z)= w_0-w_{\beta}\frac{(1+z)^{-\beta}-1}{\beta}$. In order to verify if ADD data
can provide results consistent with the widely used LD data, we also
display the constraints on the EoS parameters with the combined angular diameter distance data (SGL+CBF+BAO+WMAP9), in comparison with the luminosity distance data (SN+GRB) in Table~\ref{result}.
From the results listed, first of all, we find that the combination with angular diameter
distance data reveals no obvious disagreement between standard candle and standard ruler data, which is quite different from the previous findings \cite{Lazkoz08}.
Secondly, when discussing the tension between LD and ADD in a general framework, we note that the consistency between ADD and LD data shows up irrespective of the EoS parameterizations: there is a good match between the
universally explored CPL model and other formulations of cosmic equation of state. Thirdly, we have considered the influence
of the parameter $\beta$ in the generalized equation of state. Especially for the truncated GEoS model with $\beta=-2$, the angular diameter distance data combination reveals no tension for this EoS parametrization. These findings still hold when throwing out the Gamma-Ray burst observations and substituting WMAP9 with Planck data.

Finally, complementary conclusions are obtained from the statistical analysis of the generalized equation of state. Compared with the previous works, the combined ADD data could effectively provide
more stringent constraints on the EoS parameters $w_0$, $w_{\beta}$ and $\beta$. In this aspect, the constraint results indicates that dark energy seems to act a subdominant component at early times of the universe.
Moreover, compared with other two-parameter and time-dependent EoS parameterizations discussed in this paper, the CPL parametrization is a more favorable model to characterize the cosmic equation of state with $\beta\simeq 1$, since the combined SGL+CBF+BAO+WMAP9 data provide the best-fit value $\beta=0.751^{+0.465}_{-0.480}$. However, this conclusion still needs to be checked by future observational data of high accuracy \cite{Cao12b,Chen13,Sereno13}, which can hopefully provide significantly more restrictive constraints on cosmological parameters.

In conclusion, finding out whether the angular diameter distances and the luminosity distances are consistent is a permanent pursuit, especially
in the constraints of dark energy equation of state. Our results demonstrate that, with more general categories of standard ruler data, the controversial constraint tension between ADD and LD data does not persist at $1\sigma$. In this sense our results support the validity of the Etherington duality principle, but of course it is not a strong test because our analysis has different goals in focus.
Better understanding of the systematic uncertainties of all data used in this paper still needs to be improved.
We hope that future dark energy measurements from space \citep{Cimatti09,Laureijs11,Green12}, which may dramatically minimize systematic uncertainties by design will shed much more light into the dark universe.

\section*{Acknowledgments}

We acknowledge fruitful discussions with M. Biesiada and Z.X. Li. We thank the anonymous referee for valuable comments which helped us to improve the paper. This work was supported by the Ministry of Science and Technology National Basic Science Program (Project 973) under Grants Nos. 2012CB821804 and 2014CB845806, the Strategic Priority Research Program "The Emergence of Cosmological Structure"  of the Chinese Academy of Sciences (No. XDB09000000), the National Natural Science Foundation of China under Grants Nos. 11073005 and 11373014, the Fundamental Research Funds for the Central Universities and Scientific Research Foundation of Beijing Normal University, and China Postdoctoral Science Foundation under Grant No. 2014M550642.

\end{document}